\documentclass[prd,aps,preprintnumbers,floats,floatfix,superscriptaddress,preprintnumbers,eqsecnum,nofootinbib,notitlepage,twocolumn]{revtex4-1}
\usepackage{amsfonts,amssymb,amsmath}
\allowdisplaybreaks[1]

\begin{document}

\preprint{YITP-18-94, IPMU18-0140} 

\title{Stability of stealth magnetic field in de Sitter spacetime}
\author{Shinji Mukohyama}
\affiliation{Center for Gravitational Physics, Yukawa Institute for Theoretical Physics, Kyoto University, 606-8502, Kyoto, Japan}
\affiliation{Kavli Institute for the Physics and Mathematics of the Universe (WPI), The University of Tokyo Institutes for Advanced Study, The University of Tokyo, Kashiwa, Chiba 277-8583, Japan}

\date{\today}

 \begin{abstract}
  A detailed stability analysis is presented for the de Sitter solution with a homogeneous magnetic field that was recently found in the context of a $U(1)$ gauge theory nonminimally coupled to scalar-tensor gravity. The magnetic field is ``stealth'' in the sense that the corresponding stress-energy tensor is of the form of an effective cosmological constant and thus is isotropic despite the fact that the magnetic field has a preferred spatial direction. We study the stability of the solution against linear perturbations in the subhorizon and superhorizon limits. We then present some explicit examples that satisfy all stability conditions. The stable de Sitter solution with a homogeneous magnetic field opens up a new possibility for inflationary magnetogenesis, in which magnetic fields in the Universe at all scales may originate from a classical, homogeneous magnetic field sustained during inflation. 
\end{abstract}

\maketitle

\section{Introduction}

The origin of magnetic fields in the Universe at various scales is one of the mysteries in modern cosmology. There thus have been a large number of attempts to find a theoretical framework in which magnetic fields in the Universe are generated in the course of the cosmic history (see \cite{Widrow:2002ud,Kandus:2010nw,Durrer:2013pga,Subramanian:2015lua} for reviews). The author recently proposed a new scenario of inflationary magnetogenesis in which magnetic fields in the Universe at all scales may originate from a classical, homogeneous magnetic field sustained during inflation~\cite{Mukohyama:2016npi}. It was found that a $U(1)$ gauge theory nonminimally coupled to scalar-tensor gravity admits a cosmological attractor solution that represents a de Sitter universe with a homogeneous magnetic field, fully taking into account the backreaction of the magnetic field on the geometry.

In the standard Einstein-Maxwell theory, a homogeneous magnetic field would introduce anisotropies in the geometry through the stress-energy tensor, making it impossible to admit a de Sitter solution. One thus needs to modify the standard Einstein-Maxwell theory one way or another to make a homogeneous magnetic field and an isotropic expansion of the universe consistent with each other. In the solution found in \cite{Mukohyama:2016npi}, a homogeneous magnetic field has a preferred spatial direction but its stress-energy tensor is isotropic and is of the form of an effective cosmological constant. In this sense the magnetic field in the solution is ``stealth''.

The basic idea behind the model of \cite{Mukohyama:2016npi} is as follows. For simplicity let us first consider an action for a $U(1)$ gauge field $A_{\mu}$ of the form
\begin{equation}
 I_w = \int d^4x \sqrt{-g}\, f(w)\,, \quad w = -\frac{1}{4}F_{\mu\nu}F^{\mu\nu}\,,
\end{equation}
where $F_{\mu\nu}=\partial_{\mu}A_{\nu}-\partial_{\nu}A_{\mu}$ is the field strength and $g_{\mu\nu}$ is the spacetime metric. By taking the variation of the action with respect to $g_{\mu\nu}$, one obtains the corresponding stress-energy tensor as
\begin{equation}
 T_w^{\mu\nu} = \frac{2}{\sqrt{-g}}\frac{\delta I_w}{\delta g_{\mu\nu}} = f'(w)F^{\mu}_{\ \rho}F^{\nu\rho} + f(w)g^{\mu\nu}\,.
\end{equation}
If $f'(w)=0$ then the stress-energy tensor is proportional to $g^{\mu\nu}$ and thus admits an exact de Sitter solution. For example, this would be the case if $f(w) = f_0 + f_2 (w-w_0)^2$ with $f_0$, $f_2$ and $w_0$ constant and if $w=w_0$. However, in Friedmann-Lema\^{i}tre-Robertson-Walker (FLRW) backgrounds (including the de Sitter spacetime), $w$ for a homogeneous magnetic field decays as $1/a^4$, where $a$ is the scale factor, and thus the only constant value of $w$ that is consistent with the expansion of the universe is zero. For this reason, this simple action does not work. Let us next consider an action of the form
\begin{equation}
 I_W = \int d^4x \sqrt{-g}\, f(W)\,, \quad W = -\frac{1}{4}e^{2\phi}F_{\mu\nu}F^{\mu\nu}\,,
  \label{eqn:simpleaction}
\end{equation}
where $\phi$ is a scalar field and it is understood that some kinetic terms for $\phi$ are also added to the action. In this case the corresponding stress-energy tensor is again proportional to $g^{\mu\nu}$ if $f'(W)=0$. The main difference from the previous case is that $W$ for a homogeneous magnetic field is proportional to $e^{2\phi}/a^4$ in FLRW backgrounds and thus is constant if $e^{\phi}\propto a^2$. This is the basic idea behind the model of \cite{Mukohyama:2016npi}.

The model action of \cite{Mukohyama:2016npi} however includes not only a nonlinear function of $W$ but also other terms such as the Horndeski's nonminimal vector coupling and the Horndeski scalar terms. This is because, while the simple action (\ref{eqn:simpleaction}) supplemented by kinetic terms for $\phi$ in principle admits a de Sitter solution with a homogeneous magnetic field, the stability of such a solution requires inclusion of additional terms in the action.

The purpose of the present paper is to show a detailed stability analysis for the solution representing a de Sitter spacetime with a stealth homogeneous magnetic field. In particular we study the stability of the solution against linear perturbations in the subhorizon and superhorizon limits. We show that the stability of the solution requires inclusion of the Horndeski's nonminimal vector coupling as well as the Horndeski scalar terms. We then present some explicit examples that satisfy all stability conditions.

The rest of the present paper is organized as follows. In section~\ref{sec:model} we briefly review the model and the solutions studied in \cite{Mukohyama:2016npi}, focusing on those without the electric field. In section~\ref{sec:linearperturbation} we formulate the linear perturbation analysis and study the stability of the de Sitter solution with the stealth magnetic field against subhorizon perturbations. In section~\ref{sec:longwavelengthperturbations} we then study the stability of the solution against superhorizon perturbations. In section~\ref{sec:examples}, after showing a couple of classes of models that are unstable, we present explicit models that satisfy all stability conditions. Section~\ref{sec:summary} is then devoted to a summary of the paper and some discussions. Appendices~\ref{app:submatrices}-\ref{app:calA123} show explicit expressions of some matrices and coefficients.

\section{Review of the model}
\label{sec:model}

In this section we briefly review the model and the de Sitter solution with a homogeneous magnetic field studied in \cite{Mukohyama:2016npi}. 

\subsection{Action}

The model consists of a metric $g_{\mu\nu}$, a $U(1)$ gauge field $A_{\mu}$ and a scalar field $\phi$, described by the action 
\begin{equation}
 I = \int d^4x \sqrt{-g}
  \left[ L + L_3 + L_4 + L_5 + L_{\rm H}\right]\,, \label{eqn:action}
\end{equation}
where $L=L(X,W,Y,Z)$ is an arbitrary function of
\begin{align}
 &
  X \equiv  -\frac{1}{2}g^{\mu\nu}\partial_{\mu}\phi\partial_{\nu}\phi\,,\quad
  W \equiv  -\frac{1}{4}\mathcal{F}_{\mu\nu}\mathcal{F}^{\mu\nu}\,,\nonumber\\
 &
  Y \equiv  \mathcal{F}_{\mu\nu}\tilde{\mathcal{F}}^{\mu\nu}\,,\quad
 Z \equiv  \mathcal{F}^{\rho\mu}\mathcal{F}_{\rho}^{\ \nu}\partial_{\mu}\phi\partial_{\nu}\phi\,; 
\end{align}
$\mathcal{F}_{\mu\nu}$ and $\tilde{\mathcal{F}}^{\mu\nu}$ are defined by 
\begin{align}
& \mathcal{F}_{\mu\nu} \equiv e^{\phi}F_{\mu\nu}\,,
  \quad
  \tilde{\mathcal{F}}^{\mu\nu} \equiv e^{\phi}\tilde{F}^{\mu\nu}\,, \nonumber\\
& 
 F_{\mu\nu} \equiv \partial_{\mu}A_{\nu} - \partial_{\nu}A_{\mu}\,, \quad
  \tilde{F}^{\mu\nu} \equiv \frac{1}{2}\epsilon^{\mu\nu\rho\sigma}F_{\rho\sigma}\,,
\end{align}
and $\epsilon^{0123}=-1/\sqrt{-g}$; 
\begin{align}
 L_3 =& -G_3(X)\Box\phi\,, \nonumber\\
 L_4 =& G_4(X) R + G_{4X}(X)\left[(\Box\phi)^2-(\nabla^{\mu}\nabla_{\nu}\phi)(\nabla^{\nu}\nabla_{\mu}\phi)\right]\,,\nonumber\\
 L_5 =& G_5(X)G^{\mu\nu}\nabla_{\mu}\nabla_{\nu}\phi 
  - \frac{1}{6}G_{5X}(X) 
  \left[ (\Box\phi)^3 \right.\nonumber\\
&  - 3(\Box\phi)(\nabla^{\mu}\nabla_{\nu}\phi)(\nabla^{\nu}\nabla_{\mu}\phi) \nonumber\\
& \left.+ 2(\nabla^{\mu}\nabla_{\nu}\phi)(\nabla^{\nu}\nabla_{\rho}\phi)(\nabla^{\rho}\nabla_{\mu}\phi) \right]\,,
\end{align}
are shift-symmetric Horndeski scalar terms~\cite{Horndeski:1974wa,Deffayet:2011gz}; and
\begin{equation}
 L_{\rm H} = \xi\tilde{\mathcal{F}}^{\mu\nu}\tilde{\mathcal{F}}^{\rho\sigma}R_{\mu\nu\rho\sigma}\,, 
\end{equation}
is a simple modification ($\tilde{F}^{\mu\nu} \to \tilde{\mathcal{F}}^{\mu\nu}$) of the Horndeski's nonminimal coupling of the $U(1)$ gauge field to the Riemann tensor $R^{\mu}_{\ \nu\rho\sigma}$ of the metric $g_{\mu\nu}$~\cite{Horndeski:1976gi}. Here, the scalar field $\phi$ is normalized so that its mass dimension is zero, $G_{3,4,5}(X)$ are arbitrary functions of $X$, the subscript $X$ denotes derivative with respect to $X$, and $\xi$ is an arbitrary constant.

The action is diffeomorphism invariant and enjoys the $U(1)$ gauge symmetry
\begin{equation}
 A_{\mu} \to A_{\mu} + \partial_{\mu} \lambda\,,  
\end{equation}
where $\lambda$ is an arbitrary function. Furthermore, the action respects the global symmetry
\begin{equation}
 \phi \to \phi + \phi_0\,, \quad A_{\mu} \to e^{-\phi_0} A_{\mu}\,,
\end{equation}
where $\phi_0$ is an arbitrary constant. We assume that the function $L(X,W,Y,Z)$ is even with respect to $Y$
\begin{equation}
 L(X,W,Y,Z) = L(X,W,-Y,Z)\,. \label{eqn:Parity}
\end{equation}
This assumption ensures that the equations of motion admit a solution without the electric field~\cite{Mukohyama:2016npi}.

\subsection{Bianchi I solution with magnetic field}
\label{subsec:ansatz}

We first consider a Bianchi I spacetime 
\begin{align}
g_{\mu\nu}dx^{\mu}dx^{\nu} =& -N(t)^2dt^2 + a(t)^2
  \left[ e^{4\sigma(t)}dx^2\right. \nonumber\\
& \left. + e^{-2\sigma(t)}(dy^2+dz^2)\right]\,,
\end{align}
with
\begin{equation}
 \frac{\dot{a}}{Na} = const. \equiv H_0\,, \quad  \frac{\dot{\sigma}}{N} = const. \equiv \Sigma_0\,,
\end{equation}
where an over-dot represents derivative with respect to $t$. We further assume that the scalar field is homogeneous, $\phi = \phi(t)$, and has a constant ``velocity'' as
\begin{equation}
 \frac{\dot{\phi}}{N} = const. \,,
\end{equation}
and that the $U(1)$ gauge field represents a homogeneous magnetic field as
\begin{equation}
 A_t = A_x = 0\,,\quad 
 A_y = \frac{1}{2}Bz\,,\quad  A_z = -\frac{1}{2}By\,,  \label{eqn:ansatz-Amu}
\end{equation}
where $B$ is a constant.

Assuming that not only $H_0$, $\Sigma_0$, $B$ and $\dot{\phi}/N$ but also $\chi\equiv e^{\phi}/a^2$ are constant (see the sentences just after (\ref{eqn:simpleaction}) for the motivation for this extra assumption) and rescaling the spatial coordinates so that $\chi=1$, the equations of motion are reduced to
\begin{equation}
 0 = bL_Y\,, \label{eqn:LY=0}
\end{equation}
and
\begin{align}
 0 =& 16(1-s)^6H_0^6G_{5X}-6G_4(1-s)^2 H_0^2 - L\nonumber\\
 &  + 4\left[4\xi b^2+6(1-s)^2G_{4X}\right](1-s)^2H_0^4\,, \nonumber\\
 0 =& 8G_{5XX}(1+2s)(1-s)^6H_0^6\nonumber\\
 & + 6(1-s)^4\left[ 4(1+s)G_{4XX}+(1+4s)G_{5X}\right]H_0^4\nonumber\\
 & + 2\left\{-(1-4s)\xi b^2 + 3(1-s)^2[G_{3X} \right.\nonumber\\
 & \left.+(1+3s)G_{4X}]\right\}H_0^2-3sG_4 + (1-s)L_X \,,\nonumber\\
 0 =& 4\left[18s (1-s)^2G_{4X} -\xi(5-4s+8s^2)b^2 \right]H_0^2 \nonumber\\
 & + 72G_{5X}(1-s)^4sH_0^4 -b^2L_W - 18sG_4 \,, \label{eqn-Bianchi}
\end{align}
where $b\equiv B/H_0$ and $s\equiv\Sigma_0/H_0$. Hereafter, subscripts $W$, $Y$ and $Z$ represent derivatives with respect to them. Since $Y=0$ for the ansatz (\ref{eqn:ansatz-Amu}) and we have assumed (\ref{eqn:Parity}), the first equation (\ref{eqn:LY=0}) is automatically satisfied. Therefore we have three algebraic equations (\ref{eqn-Bianchi}) to be solved with respect to the three parameters ($H_0$, $b$, $s$) of the ansatz. Generically they admit a solution (or a set of solutions).

\subsection{de Sitter solution with magnetic field}
\label{subsec:dSnoE}

By fine-tuning one of parameters in the action, one can then take the limit $s\to 0$ so that the spacetime becomes a de Sitter,
\begin{equation}
g_{\mu\nu}dx^{\mu}dx^{\nu} = -N(t)^2dt^2 + a(t)^2(dx^2+dy^2+dz^2)\,. 
\end{equation}
Then, under the assumption (\ref{eqn:Parity}), the independent equations of motion are 
\begin{align}
0 =&   16G_{5X}H_0^6 + 8(2\xi b^2+3G_{4X})H_0^4 -6G_4H_0^2 - L\,,\nonumber\\
 0 =&   8G_{5XX}H_0^6 + 6(4G_{4XX}+G_{5X})H_0^4 \nonumber\\
 & + 2\left[ -\xi b^2 + 3(G_{3X}+G_{4X})\right]H_0^2+ L_X\,,\nonumber\\
0 =&   20\xi H_0^2 + L_W\,. \label{eqn:dSnoE}
\end{align}
These are three algebraic equations. One of the three equations (or one combination of them) simply represents the fine-tuning of one of parameters in the action and the remaining two equations can generically be solved with respect to the two parameters ($H_0$, $b$) of the ansatz. 

If we relax the fine-tuning without abandoning the discrete symmetry (\ref{eqn:Parity}) then the solution goes back to the Bianchi I solution with the three parameters ($H_0$, $b$, $s$) described in the previous subsection.

\subsection{Attractor condition}
\label{subsec:attractor}

Ref.~\cite{Mukohyama:2016npi} found the necessary and sufficient condition under which the de Sitter solution with a homogeneous magnetic field introduced in subsection~\ref{subsec:dSnoE} is stable against homogeneous linear perturbations, under the assumption (\ref{eqn:Parity}). The stability condition, or the attractor condition, is $\mathcal{A}/\mathcal{N}>0$, where 
\begin{align}
 \mathcal{N} =& 2\zeta_3g_h(\zeta_3-8\zeta_1)b^2+\zeta_1(\zeta_1\zeta_2+3\zeta_3^2)\,,\nonumber\\
 \mathcal{A} =& 56b^6g_h^3-4(9\zeta_1+\zeta_2+15\zeta_3)g_h^2b^4-2g_h\zeta_4(\zeta_1-\zeta_3)b^3\nonumber\\
 & 
  +\left[6(-\zeta_1^2+\zeta_1\zeta_2+2\zeta_1\zeta_3+2\zeta_3^2)g_h\right.\nonumber\\
 & \left.+\zeta_5(\zeta_1-\zeta_3)^2\right]b^2+\frac{3}{2}\zeta_1\zeta_4(\zeta_1-\zeta_3)b \,,
 \label{eqn:def-calN-calA}
\end{align}
$g_h$ and $\zeta_i$ ($i=1,\cdots 5$) are constants defined by
\begin{align}
 g_h =& \xi H_0^2/M_{\rm Pl}^2\,, \quad
  \zeta_1 = 2b^2g_h+g_4-4g_{4x}-4g_{5x}\,,\nonumber\\
 \zeta_2 =& 2b^2g_h+6g_{3x}+24g_{3xx}+72g_{4xx}+96g_{4xxx}\nonumber\\
& +6g_{5x}+48g_{5xx}+32g_{5xxx}+4l_{xx}  \,,\nonumber\\
 \zeta_3 =& 4b^2g_h+2g_{3x}+4g_{4x}+16g_{4xx}+6g_{5x}+8g_{5xx}\,,\nonumber\\
 \zeta_4 =& -4(g_h+l_{xw})b  \,,\quad 
  \zeta_5 = -12g_h-b^2l_{ww}\,, \label{eqn:def-gh-zeta}
\end{align}
and 
\begin{align}
 &
  L_{XX} = l_{xx}\frac{M_{\rm Pl}^2}{H_0^2}\,, \ 
  L_{XW} = l_{xw}\frac{M_{\rm Pl}^2}{H_0^2}\,, \ 
  L_{WW} = l_{ww}\frac{M_{\rm Pl}^2}{H_0^2}\,, \nonumber\\
 &
  G_{3X} = g_{3x}\frac{M_{\rm Pl}^2}{H_0^2}\,, \ 
  G_{3XX} = g_{3xx}\frac{M_{\rm Pl}^2}{H_0^4}\,, \nonumber\\
 &
  G_4 = g_4M_{\rm Pl}^2\,, \ 
  G_{4X} = g_{4x}\frac{M_{\rm Pl}^2}{H_0^2}\,, \
  G_{4XX} = g_{4xx}\frac{M_{\rm Pl}^2}{H_0^4}\,, \nonumber\\
 & 
  G_{4XXX} = g_{4xxx}\frac{M_{\rm Pl}^2}{H_0^6}\,,\ 
  G_{5X} = g_{5x}\frac{M_{\rm Pl}^2}{H_0^4}\,, \nonumber\\
 & 
  G_{5XX} = g_{5xx}\frac{M_{\rm Pl}^2}{H_0^6}\,, \
  G_{5XXX} = g_{5xxx}\frac{M_{\rm Pl}^2}{H_0^8}\,. \label{eqn:dimensionlessparameters}
\end{align}
It is understood that the left hand sides of (\ref{eqn:dimensionlessparameters}) are evaluated at the de Sitter solution with a homogeneous magnetic field and thus are constant.

We shall later see that the absence of ghosts implies that $\mathcal{N}>0$. Therefore the attractor condition is
\begin{equation}
 \mathcal{A}>0\,. \label{eqn:attractor}
\end{equation}

\section{Linear perturbation}
\label{sec:linearperturbation}

 In this section we consider inhomogeneous linear perturbations around the de Sitter attractor solution with a homogeneous magnetic field ($H_0>0$, $s=0$, $b\ne 0$) and seek the no-ghost conditions and the sound speeds in the subhorizon limit.

 \subsection{Decomposition of linear perturbation}

 Around the general background introduced in subsection \ref{subsec:ansatz} we introduce the components $\Phi$, $V_i$ and $h_{ij}$ ($i,j=1,2,3$) of the metric perturbation $\delta g_{\mu\nu}$, the components $\delta A_a$ ($a=0,\cdots,3$) of the perturbation of the vector field $\delta A_{\mu}$ and the perturbation $\pi$ of the scalar field $\phi$ as
\begin{align}
&  \delta g_{\mu\nu}dx^{\mu}dx^{\nu} =
   -2\Phi\, e^0e^0 + V_i\, e^0e^i + V_i\, e^ie^0 + h_{ij}\, e^ie^j\,, \nonumber\\
  &  \delta A_{\mu}dx^{\mu} = \delta A_a e^a\,,\quad 
   \delta\phi = \pi\,, 
 \end{align}
 where 
\begin{align}
& e^0 = N(t)dt\,, \quad e^1 = a(t) e^{2\sigma(t)}dx\,, \nonumber\\
&  e^2 = a(t) e^{-\sigma(t)}dy\,, \quad   e^3 = a(t) e^{-\sigma(t)}dz\,,
\end{align}
and the Einstein's summation rule is employed. We then decompose each component of the perturbation into spatial Fourier modes. For this purpose, we first introduce scalar harmonics $Y_n$ ($n = cc, cs, sc, ss$) as
\begin{align}
  Y_{cc} =& \cos(k_x x)\cos(k_y y + k_z z)\,,\nonumber\\
  Y_{cs} =& \cos(k_x x)\sin(k_y y + k_z z)\,,\nonumber\\
  Y_{sc} =& \sin(k_x x)\cos(k_y y + k_z z)\,,\nonumber\\
  Y_{ss} =& \sin(k_x x)\sin(k_y y + k_z z)\,, 
\end{align}
and then define odd and even vector harmonics $Y^{\rm odd/even}_{n, p}$ ($n = cc, cs, sc, ss$; $p, q = 2, 3$) as
\begin{equation}
  Y^{\rm odd}_{n, p} = \epsilon_{pq}\frac{\partial}{\partial x^q}Y_n\,, \quad 
  Y^{\rm even}_{n, p} =\frac{\partial}{\partial x^p}Y_n\,, 
\end{equation}
where $x^2=y$, $x^3=z$, and $\epsilon_{pq}$ is the two-dimensional Levi-Civita symbol ($\epsilon_{23}=-\epsilon_{32}=1$ and $\epsilon_{22}=\epsilon_{33}=0$), and odd and even tensor harmonics $Y^{\rm odd/even}_{n, pq}$ ($n = cc, cs, sc, ss$; $p, q = 2, 3$) as
\begin{equation}
  Y^{\rm odd/even}_{n, pq} =\frac{1}{2}\left(\frac{\partial}{\partial x^p}Y^{\rm odd/even}_{n,q}+\frac{\partial}{\partial x^q} Y^{\rm odd/even}_{n,p}\right)\,.
\end{equation}
The components of the perturbations are then decomposed as
\begin{align}
& \pi = \int\frac{d^3k}{(2\pi)^3} \pi_n\, Y_n\,,\quad
 \Phi = \int\frac{d^3k}{(2\pi)^3} \Phi_n\, Y_n\,,\nonumber\\
&
  V_1 = \int\frac{d^3k}{(2\pi)^3} \chi_n\, Y_n\,,\quad
  h_{11} = \int\frac{d^3k}{(2\pi)^3}\psi_n\, Y_n\,,\nonumber\\
&
   \delta A_0 = \int\frac{d^3k}{(2\pi)^3} A_{0,n}\, Y_n\,,\quad
   \delta A_1 = \int\frac{d^3k}{(2\pi)^3} A_{1,n}\, Y_n\,,\nonumber\\
&  V_p = \int\frac{d^3k}{(2\pi)^3} (V^{\rm odd}_nY^{\rm odd}_{n,p}+V^{\rm even}_nY^{\rm even}_{n,p})\,,\nonumber\\
 &
   h_{1p} = h_{p1} = \int\frac{d^3k}{(2\pi)^3} (h^{\rm odd}_{1,n}Y^{\rm odd}_{n,p}+\beta_nY^{\rm even}_{n,p})\,,\nonumber\\
& \delta A_p = \int\frac{d^3k}{(2\pi)^3} (A^{\rm odd}_nY^{\rm odd}_{n,p}+A^{\rm even}_nY^{\rm even}_{n,p})\,, \nonumber\\
&
 h_{pq} = \int\frac{d^3k}{(2\pi)^3} (h^{\rm odd}_nY^{\rm odd}_{n,pq}+E_nY^{\rm even}_{n,pq}+\tau_nY_n\delta_{pq})\,, 
\end{align}
where $p, q = 2, 3$ and the Einstein's summation rule for $n = cc, cs, sc, ss$ is understood.

We fix gauge freedom associated with the spacetime coordinate transformation and the $U(1)$ gauge transformation as
\begin{equation}
 \beta_n=h^{\rm odd}_n = E_n=\tau_n= 0\,,\quad
  A^{\rm even}_n=0\,.
\end{equation}

\subsection{General quadratic action}

Substituting the decomposition of perturbations to the action, imposing the gauge condition and expanding the action up to the quadratic order in perturbations, we obtain the quadratic action for the linear perturbations around a general background introduced in subsection \ref{subsec:ansatz} as
\begin{align}
 I^{(2)} = & \frac{1}{2} \int\frac{d^3k}{(2\pi)^3} \int Na^3dt
  \left[ \frac{1}{N^2}\dot{\bf Y}_n^{\intercal} {\bf K}\dot{\bf Y}_n
   + \frac{1}{N}\left(\dot{{\bf Y}}_n^{\intercal}{\bf M} {\bf Y}_n \right.\right.\nonumber\\
 & \left.
 + {\bf Y}_n^{\intercal}{\bf M}^{\intercal}\dot{{\bf Y}}_n\right)
 - {\bf Y}_n^{\intercal}{\bf V}{\bf Y}_n + \left({\bf Z}_n^{\intercal}{\bf A} {\bf Y}_n + {\bf Y}_n^{\intercal}{\bf A}^{\intercal} {\bf Z}_n\right) \nonumber\\
 & \left. + \frac{1}{N}\left({\bf Z}_n^{\intercal}{\bf B}\dot{{\bf Y}}_n+\dot{{\bf Y}}_n^{\intercal}{\bf B}^{\intercal}{\bf Z}_n\right) + {\bf Z}_n^{\intercal}{\bf C}{\bf Z}_n  \right]\,,
  \label{eqn:generalI2}
\end{align}
where the superscript ${\intercal}$ represents the transpose operation, 
\begin{equation}
 {\bf Y}_n  =
  \left(
   \begin{array}{c}
    k_{\perp}a^2A^{\rm odd}_n \\
    k_{\perp}h^{\rm odd}_{1,n}\\
    a^2A_{1,n}\\
    \pi_n\\
    \psi_n
   \end{array}
		 \right),
\end{equation}
represent dynamical degrees of freedom, 
\begin{align}
&
 {\bf Z}_{cc}  = 
  \left(
   \begin{array}{c}
    k_{\perp}V^{\rm odd}_{sc}\\
    k_{\perp}V^{\rm even}_{cc}\\
    a^2A_{0,sc}\\
    \chi_{sc}\\
    \frac{a}{k_{\perp}}H_0\Phi_{cc}\\
   \end{array}
		 \right),\, 
 {\bf Z}_{cs}  = 
  \left(
   \begin{array}{c}
    k_{\perp}V^{\rm odd}_{ss}\\
    k_{\perp}V^{\rm even}_{cs}\\
    a^2A_{0,ss}\\
    \chi_{ss}\\
    \frac{a}{k_{\perp}}H_0\Phi_{cs}\\
   \end{array}
		 \right),\nonumber\\
&
 {\bf Z}_{sc}  = 
  \left(
   \begin{array}{c}
    -k_{\perp}V^{\rm odd}_{cc}\\
    k_{\perp}V^{\rm even}_{sc}\\
    -a^2A_{0,cc}\\
    -\chi_{cc}\\
    \frac{a}{k_{\perp}}H_0\Phi_{sc}\\
   \end{array}
		 \right),\, 
 {\bf Z}_{ss}  = 
  \left(
   \begin{array}{c}
    -k_{\perp}V^{\rm odd}_{cs}\\
    k_{\perp}V^{\rm even}_{ss}\\
    -a^2A_{0,cs}\\
    -\chi_{cs}\\
    \frac{a}{k_{\perp}}H_0\Phi_{ss}\\
   \end{array}
		 \right),
\end{align}
represent non-dynamical degrees of freedom, $k_{\perp} = \sqrt{(k_y)^2 + (k_z)^2}$ and the Einstein's summation rule for $n = cc, cs, sc, ss$ is again understood. Important points are that the matrices ${\bf K}$, ${\bf M}$, ${\bf V}$, ${\bf A}$, ${\bf B}$ and ${\bf C}$ are common for all $n = cc, cs, sc, ss$ and that there are no coupling between ($Y_n$, $Z_n$) and ($Y_{n'}$, $Z_{n'}$) with $n \ne n'$. In general, however, even and odd sectors within ($Y_n$, $Z_n$) couple with each other.

Integrating out the non-dynamical variables ${\bf Z}_n$, one obtains the quadratic action for the dynamical variables ${\bf Y}_n$ as
\begin{align}
 \tilde{I}^{(2)} =& \frac{1}{2} \int\frac{d^3k}{(2\pi)^3} \int Na^3dt
  \left[ \frac{1}{N^2}\dot{\bf Y}_n^{\intercal} \bar{\bf K}\dot{\bf Y}_n
   + \frac{1}{N}\left(\dot{{\bf Y}}_n^{\intercal}\bar{\bf M} {\bf Y}_n \right.\right.\nonumber\\
& \left.\left.
 + {\bf Y}_n^{\intercal}\bar{\bf M}^{\intercal}\dot{{\bf Y}}_n\right)
 - {\bf Y}_n^{\intercal}\bar{\bf V}{\bf Y}_n \right]\,,
  \label{eqn:reducedaction}
\end{align}
where
\begin{align}
& \bar{\bf K} = {\bf K} - {\bf B}^{\intercal}{\bf C}^{-1}{\bf B}\,,\quad
 \bar{\bf M} = \frac{1}{2}\left(\bar{\bf m}-\bar{\bf m}^{\intercal}\right)\,,\nonumber\\
 & \bar{\bf V} = {\bf V} + {\bf A}^{\intercal}{\bf C}^{-1}{\bf A} + \frac{1}{2Na^3}\partial_t\left[a^3\left(\bar{\bf m}+\bar{\bf m}^{\intercal}\right)\right]\,,
\end{align}
and
\begin{equation}
 \bar{\bf m} = {\bf M} - {\bf B}^{\intercal}{\bf C}^{-1}{\bf A}\,,\label{eqn:reducedmatrices}
\end{equation}

It is then straightforward to diagonalize the kinetic matrix $\bar{\bf K}$ as
 \begin{equation}
  \bar{\bf K} \to \tilde{\bf K} = {\bf U}^{\intercal}\bar{\bf K}{\bf U}\,,\quad
   {\bf U} = {\bf U}_1{\bf U}_2{\bf U}_3{\bf U}_4{\bf U}_5\,,
   \label{eqn:tildeK}
 \end{equation}
 where 
\begin{align}
 {\bf U}_1 =&  \left(\begin{array}{ccccc}
1 & -\frac{\bar{K}_{12}}{\bar{K}_{11}} & -\frac{\bar{K}_{13}}{\bar{K}_{11}} & -\frac{\bar{K}_{14}}{\bar{K}_{11}} & -\frac{\bar{K}_{15}}{\bar{K}_{11}} \\
0 & 1 & 0 & 0 & 0\\
0 & 0 & 1 & 0 & 0\\
0 & 0 & 0 & 1 & 0\\
0 & 0 & 0 & 0 & 1\\
\end{array}
			  \right)\,, \nonumber\\
 {\bf U}_2 =&  \left(\begin{array}{ccccc}
1 & 0 & 0 & 0 & 0\\
0 & 1 & -\frac{\bar{K}_{23}}{\bar{K}_{22}} & -\frac{\bar{K}_{24}}{\bar{K}_{22}} & -\frac{\bar{K}_{25}}{\bar{K}_{22}} \\
0 & 0 & 1 & 0 & 0\\
0 & 0 & 0 & 1 & 0\\
0 & 0 & 0 & 0 & 1\\
\end{array}
			  \right)\,, \nonumber\\
 {\bf U}_3 =&  \left(\begin{array}{ccccc}
1 & 0 & 0 & 0 & 0\\
0 & 1 & 0 & 0 & 0\\
0 & 0 & 1 & -\frac{\bar{K}_{34}}{\bar{K}_{33}} & -\frac{\bar{K}_{35}}{\bar{K}_{33}} \\
0 & 0 & 0 & 1 & 0\\
0 & 0 & 0 & 0 & 1\\
\end{array}
			  \right)\,, \nonumber\\
 {\bf U}_4 =&  \left(\begin{array}{ccccc}
1 & 0 & 0 & 0 & 0\\
0 & 1 & 0 & 0 & 0\\
0 & 0 & 1 & 0 & 0\\
0 & 0 & 0 & 1 & -\frac{\bar{K}_{45}}{\bar{K}_{44}} \\
0 & 0 & 0 & 0 & 1\\
\end{array}
			  \right)\,,
\end{align}
and ${\bf U}_5={\rm diag}(1,k/k_{\perp},k/k_{\perp},1,k^2/k^2_{\perp})$. Here $k=\sqrt{(k_x)^2+(k_{\perp})^2}$. Correspondingly, the friction matrix $\bar{\bf M}$ and the mass matrix $\bar{\bf V}$ transform as 
\begin{align}
\bar{\bf M} \to& \tilde{\bf M} 
= \frac{1}{2}\left(\tilde{\bf m}-\tilde{\bf m}^{\intercal}\right)\,,\nonumber\\
\bar{\bf V} \to& \tilde{\bf V} = {\bf U}^{\intercal}\bar{\bf V}{\bf U} - \frac{1}{N}\left(\dot{{\bf U}}^{\intercal}\bar{\bf M}{\bf U}+{\bf U}^{\intercal}\bar{\bf M}^{\intercal}\dot{{\bf U}}\right)\nonumber\\
 & - \frac{1}{N^2}\dot{{\bf U}}^{\intercal}\bar{\bf K}\dot{{\bf U}} + \frac{1}{2Na^3}\partial_t\left[a^3\left(\tilde{\bf m}+\tilde{\bf m}^{\intercal}\right)\right]\,,
\end{align}
where
\begin{equation}
 \tilde{\bf m} = {\bf U}^{\intercal}\bar{\bf M}{\bf U} + \frac{1}{N}{\bf U}^{\intercal}\bar{\bf K}\dot{{\bf U}}\,.\label{eqn:def-mtilde}
\end{equation}
The quadratic action is now written as
\begin{align}
 \tilde{I}^{(2)} = & \frac{1}{2} \int\frac{d^3k}{(2\pi)^3} \int Na^3dt
  \left[ \frac{1}{N^2}\dot{\tilde{\bf Y}}_n^{\intercal} \tilde{\bf K}\dot{\tilde{\bf Y}}_n \right.\nonumber\\
& \left.
   + \frac{1}{N}\left(\dot{{\tilde{\bf Y}}}_n^{\intercal}\tilde{\bf M} \tilde{\bf Y}_n + \tilde{\bf Y}_n^{\intercal}\tilde{\bf M}^{\intercal}\dot{{\tilde{\bf Y}}}_n\right)
 - \tilde{\bf Y}_n^{\intercal}\tilde{\bf V}\tilde{\bf Y}_n \right]\,,
  \label{eqn:reducedaction-diagonal}
\end{align}
where the new variables $\tilde{\bf Y}_n$ are related to the original variables ${\bf Y}_n$ via ${\bf Y}_n={\bf U}\tilde{\bf Y}_n$.

\subsection{Quadratic action around de Sitter background with magnetic field}

In the de Sitter background with a homogeneous magnetic field, the matrices in (\ref{eqn:generalI2}) are greatly simplified and all their components are written in terms of $p_{\perp}\equiv k_{\perp}/a$, $p_x\equiv k_x/a$, $b$, $g_h$, $g_4$ and $\zeta_{\alpha}$ ($\alpha=1,\cdots, 7$), where $g_h$ and $\zeta_{\alpha}$ are defined in (\ref{eqn:def-gh-zeta}) and
\begin{equation}
\zeta_6 = 4(-3g_h+2l_z)\,, \ 
  \zeta_7 = 4(-3g_h+2l_z+4b^2l_{yy})\,.
\end{equation}
Here, $l_z$ and $l_{yy}$ are defined as
\begin{equation}
 L_{Z} = l_{z}\frac{M_{\rm Pl}^2}{H_0^2}\,, \quad
     L_{YY} = l_{yy}\frac{M_{\rm Pl}^2}{H_0^2}\,,  \label{eqn:def-lz-lyy}
\end{equation}
where it is understood that the left hand sides of (\ref{eqn:def-lz-lyy}) are evaluated at the de Sitter solution with a homogeneous magnetic field. It is convenient to decompose the matrices into sub-matrices as 
\begin{align}
& {\bf K} = \left(\begin{array}{cc}
      {\bf K}_{\rm odd}&  {\bf 0}\\
	    {\bf 0} &  {\bf K}_{\rm even}
	     \end{array}\right),\,
 {\bf M} = \left(\begin{array}{cc}
      {\bf 0} &  {\bf M}_{\rm mix}\\
	    -{\bf M}_{\rm mix}^{\intercal}& {\bf M}_{\rm even}
	     \end{array}\right),\nonumber\\
&
 {\bf V} = \left(\begin{array}{cc}
      {\bf V}_{\rm odd}&  {\bf V}_{\rm mix}\\
	    {\bf V}_{\rm mix}^{\intercal}&  {\bf V}_{\rm even}
	     \end{array}\right),\,
{\bf A} = \left(\begin{array}{cc}
      {\bf A}_{\rm odd}&  {\bf A}_{\rm mixI}\\
	    {\bf A}_{\rm mix II}&  {\bf A}_{\rm even}
	     \end{array}\right),\nonumber\\
&
 {\bf B} = \left(\begin{array}{cc}
      {\bf B}_{\rm odd}&  {\bf 0}\\
	    {\bf B}_{\rm mix}&  {\bf B}_{\rm even}
	     \end{array}\right),\,
 {\bf C} = \left(\begin{array}{cc}
      {\bf C}_{\rm odd}&  {\bf C}_{\rm mix}\\
	    {\bf C}_{\rm mix}^{\intercal}&  {\bf C}_{\rm even}
	     \end{array}\right), \label{eqn:submatrices}
\end{align}
where explicit expressions for the sub-matrices are given in Appendix~\ref{app:submatrices}.

If $b=0$ (we have already set $s=0$) then, since $L$ is assumed to be even with respect to $Y$ in (\ref{eqn:Parity}), the theory and the background respect the parity invariance and, as a result, the mixing matrices ${\bf M}_{\rm mix}$, ${\bf V}_{\rm mix}$, ${\bf A}_{\rm mix I}$, ${\bf A}_{\rm mix II}$, ${\bf B}_{\rm mix}$ and ${\bf C}_{\rm mix}$ vanish. In general they do not vanish and thus the even and odd perturbations do mix.

The quadratic action for the dynamical variables ${\bf Y}_n$ is then given by the formulas (\ref{eqn:reducedaction})-(\ref{eqn:reducedmatrices}). After diagonalization of the kinetic matrix, one obtains (\ref{eqn:reducedaction-diagonal}) with (\ref{eqn:tildeK})-(\ref{eqn:def-mtilde}).

\subsection{Subhorizon limit}

For the theoretical consistency, one needs to analyze the stability of the background in the UV, i.e. in the subhorizon limit $p^2\gg H_0^2$, where $p=\sqrt{p_x^2+p_{\perp}^2}=k/a$. This ensures the absence of instability whose timescale is parametrically shorter than the cosmological timescale $\sim 1/H_0$. The stability of the solution against perturbations with $p^2\ll H_0^2$ shall be studied in the next section.

In the subhorizon limit $p^2\gg H_0^2$, introducing a small bookkeeping parameter $\epsilon$ so that $H_0^2/p^2=\mathcal{O}(\epsilon^2)$, components of the matrices in the quadratic action (\ref{eqn:reducedaction-diagonal}) are simplified as
 \begin{align}
 \tilde{\bf K} =&
  \left(\begin{array}{ccccc}
   \tilde{K}_1 &  0 &  0 & 0 & 0 \\
	 0 &  \tilde{K}_2 &  0 & 0 & 0 \\
	 0 &  0 &  \tilde{K}_3 & 0 & 0 \\
	 0 &  0 &  0 & \tilde{K}_4 & 0 \\
	 0 &  0 &  0 & 0 & \tilde{K}_5
	  \end{array}
 \right) + \mathcal{O}(\epsilon^2)\,, \nonumber\\
 \tilde{\bf M} =&
  \left(\begin{array}{ccccc}
   0 &  0 &  0 & \tilde{M}_1 & \tilde{M}_2\\
	 0 &  0 &  \tilde{M}_3 & 0 & 0\\
	 0 &  -\tilde{M}_3 &  0 & 0 & 0\\
	 -\tilde{M}_1 &  0 &  0 & 0 & 0\\
 	 -\tilde{M}_2 &  0 &  0 & 0 & 0
	  \end{array}
 \right) + p\mathcal{O}(\epsilon^2)\,, \nonumber\\
 \tilde{\bf V} =&
  \left(\begin{array}{ccccc}
   \tilde{V}_1 &  0 &  0 & 0 & 0 \\
	 0 &  \tilde{V}_2 &  0 & 0 & 0 \\
	 0 &  0 &  \tilde{V}_3 & 0 & 0 \\
	 0 &  0 &  0 & \tilde{V}_4 & \tilde{V}_6 \\
	 0 &  0 &  0 & \tilde{V}_6 & \tilde{V}_5
	  \end{array}
 \right)+ p^2\mathcal{O}(\epsilon^2)\,, 
\end{align}
where
 \begin{align}
\tilde{K}_1 =& \zeta_6\,,\nonumber\\
\tilde{K}_2 =& \left(\frac{p_{\perp}^2}{\zeta_1}+ \frac{p_x^2}{\zeta_1-2b^2g_h}\right)^{-1}p^2\,,\nonumber\\
\tilde{K}_3 =& \left(\frac{p_{\perp}^2}{\zeta_7}+\frac{p_x^2}{\zeta_6}\right)^{-1}p^2\,,\nonumber\\
\tilde{K}_4 =& \frac{c_1p_{\perp}^4+2c_2p_{\perp}^2p_{x}^2+\mathcal{N}p_{x}^4}{[(\zeta_1-\zeta_3)p^2+3b^2g_hp_{\perp}^2]^2}\,,\nonumber\\
\tilde{K}_5 =& \frac{(\zeta_1-2b^2g_h)\mathcal{N}p^4}{4(c_1p_{\perp}^4+2c_2p_{\perp}^2p_{x}^2+\mathcal{N}p_{x}^4)}\,,\nonumber\\
\tilde{M}_1 =& \frac{p_{\perp}(c_3p_{\perp}^2+c_4p_{x}^2)}{(\zeta_1-\zeta_3)p^2+3b^2g_hp_{\perp}^2}\,,\nonumber\\
\tilde{M}_2 =& \frac{bg_hp_{\perp}p^2(\zeta_1-2b^2g_h)(c_5p_{\perp}^2+c_6p_{x}^2)}{\zeta_1(c_1p_{\perp}^4+2c_2p_{\perp}^2p_{x}^2+\mathcal{N}p_{x}^4)}\,,\nonumber\\
\tilde{M}_3 =& -\frac{2bg_hp_{\perp}}{\zeta_1}\tilde{K}_2\,,\nonumber\\
\tilde{V}_1 =& \zeta_5p_{\perp}^2+4g_h\left(\frac{4b^2g_h}{\zeta_1}-3\right)p_{x}^2\,,\nonumber\\
\tilde{V}_2 =& g_4p^2 \,,\nonumber\\
\tilde{V}_3 =& \frac{4g_hp^2[3(2b^2g_h-\zeta_1)p_{\perp}^2+(4b^2g_h-3\zeta_1)p_{x}^2]}{(\zeta_1-2b^2g_h)p_{\perp}^2+\zeta_1p_{x}^2}\,,\nonumber\\
\tilde{V}_4 =& \frac{c_7p_{\perp}^6+c_8p_{\perp}^4p_{x}^2+c_9p_{\perp}^2p_{x}^4+c_{10}p_{x}^6}{[(\zeta_1-\zeta_3)p^2+3b^2g_hp_{\perp}^2]^2}\,,\nonumber\\
\tilde{V}_5 =& \frac{p^4(c_{11}p_{\perp}^6+c_{12}p_{\perp}^4p_{x}^2+c_{13}p_{\perp}^2p_{x}^4+c_{14}p_{x}^6)}{[c_1p_{\perp}^4+2c_2p_{\perp}^2p_{x}^2+\mathcal{N}p_{x}^4]^2}\,,\nonumber\\
\tilde{V}_6 =& \frac{b^2g_hp_{\perp}^2p^2(c_{15}p_{\perp}^4+2c_{16}p_{\perp}^2p_{x}^2+c_{17}p_{x}^4)}{[(\zeta_1-\zeta_3)p^2+3b^2g_hp_{\perp}^2][c_1p_{\perp}^4+2c_2p_{\perp}^2p_{x}^2+\mathcal{N}p_{x}^4]}\,. \label{eqn:tildeK-tildeM-tildeV}
\end{align}
Here, $\mathcal{N}$ is defined in (\ref{eqn:def-calN-calA}) and $c_1, \cdots, c_{17}$ are shown in Appendix~\ref{app:coefficients}.

\subsection{No-ghost conditions}

All components $\tilde{K}_{1,2,3,4,5}$ of the diagonalized kinetic matrix in the subhorizon limit are positive for ${}^{\forall}p_{\perp}$ and ${}^{\forall}p_x$ such that $p^2\ne 0$, if and only if $\mathcal{N}_a>0$ ($a=1,\cdots,7$), where
\begin{align}
 & \mathcal{N}_1 = \zeta_1\,,\quad \mathcal{N}_2 = \zeta_1-2b^2g_h\,,
  \nonumber\\
 &  \mathcal{N}_3 = \zeta_6\,,\quad \mathcal{N}_4=\zeta_7\,,\quad \mathcal{N}_5=\mathcal{N}\,, \label{eqn:def-calN1to5}
\end{align}
and
\begin{equation}
 \mathcal{N}_6 = c_1\,,\quad \mathcal{N}_7 = c_2 + \sqrt{c_1\mathcal{N}}\,.
\end{equation}
Here, the positivity of the last three $\mathcal{N}_{5,6,7}$ is necessary and sufficient for $c_1p_{\perp}^4+2c_2p_{\perp}^2p_{x}^2+\mathcal{N}p_{x}^4$ to be positive for ${}^{\forall}p_{\perp}$ and ${}^{\forall}p_x$ such that $p^2\ne 0$.

One can actually show that
\begin{align}
 \mathcal{N}_6 =& \frac{\mathcal{N}_2}{4\mathcal{N}_1^2}(\mathcal{N}_1-\mathcal{N}_2)^2(4\mathcal{N}_1-\zeta_3)^2+\frac{\mathcal{N}_5}{4\mathcal{N}_1^2}(\mathcal{N}_1+\mathcal{N}_2)^2\,,\nonumber\\
 \mathcal{N}_7 =& \frac{\mathcal{N}_5}{2\mathcal{N}_1}(\mathcal{N}_1+\mathcal{N}_2) + \sqrt{\mathcal{N}_5\mathcal{N}_6}\,.
\end{align}
Therefore the positivity of $\mathcal{N}_6$ and $\mathcal{N}_7$ follows from other conditions and the no-ghost condition is simply
\begin{equation}
 \mathcal{N}_a >0\,, \quad (a=1,\cdots,5)\,. \label{eqn:noghost}
\end{equation}

\subsection{Sound speeds}

The squared sound speeds $c_s^2$ of the five modes are determined by 
\begin{align}
 0 =& \lim_{\epsilon\to 0}\det \left[ c_s^2 \tilde{\bf K} -2i\sqrt{c_s^2}\tilde{\bf M}/p - \tilde{\bf V}/p^2 \right]\nonumber\\
 =& \tilde{K}_1\tilde{K}_2\tilde{K}_3\tilde{K}_4\tilde{K}_5
  \times \left[(c_s^2)^2-2\alpha_1c_s^2+\alpha_2\right]\nonumber\\
 & \times \left[(c_s^2)^3-3\alpha_3(c_s^2)^2+3\alpha_4c_s^2-\alpha_5\right]\,,
 \label{eqn:dispersionrelation}
\end{align}
where
\begin{align}
 \alpha_1 =&
  \frac{b^2\mathcal{N}_1\mathcal{N}_3\mathcal{S}_1+2\mathcal{N}_2(\mathcal{N}_1+2\mathcal{N}_2)\mathcal{S}_2}{2b^2\mathcal{N}_1\mathcal{N}_2\mathcal{N}_3}\frac{p_x^2}{p^2} \nonumber\\
 & 
  + \frac{b^2\mathcal{N}_4\mathcal{S}_1+2(\mathcal{N}_1+2\mathcal{N}_2)\mathcal{S}_2}
  {2b^2\mathcal{N}_1\mathcal{N}_4}\frac{p_{\perp}^2}{p^2}\,, \nonumber\\
 \alpha_2 =&
  \frac{2[(\mathcal{N}_1+2\mathcal{N}_2)p_x^2+3\mathcal{N}_2p_{\perp}^2](\mathcal{N}_3p_{\perp}^2+\mathcal{N}_4p_x^2)}{b^2\mathcal{N}_1\mathcal{N}_2\mathcal{N}_3\mathcal{N}_4 p^4}\mathcal{S}_1\mathcal{S}_2\,,\nonumber\\
 \alpha_3 =&
  \frac{\mathcal{N}_1\mathcal{S}_3p_x^2+\mathcal{N}_2\mathcal{S}_4p_{\perp}^2}
  {3b^2\mathcal{N}_1^2\mathcal{N}_2\mathcal{N}_3\mathcal{N}_5 p^2}\,,\nonumber\\
 \alpha_4 =&
  \frac{\mathcal{S}_5p_x^4+2(\mathcal{S}_7-\sqrt{\mathcal{S}_5\mathcal{S}_6})p_x^2p_{\perp}^2+\mathcal{S}_6p_{\perp}^4}
  {3b^2\mathcal{N}_1\mathcal{N}_2\mathcal{N}_3\mathcal{N}_5 p^4}\,,\nonumber\\
 \alpha_5 =&
  \frac{\mathcal{S}_9p_x^4+2(\mathcal{S}_{11}-\sqrt{\mathcal{S}_9\mathcal{S}_{10}})p_x^2p_{\perp}^2+\mathcal{S}_{10}p_{\perp}^4}
  {b^2\mathcal{N}_1\mathcal{N}_2\mathcal{N}_3\mathcal{N}_5 p^6}\nonumber\\
&
  \times 
  [2(\mathcal{N}_1+\mathcal{N}_2)\mathcal{S}_2p_x^2+b^2\mathcal{N}_1\mathcal{S}_8p_{\perp}^2]\,,\label{eqn:def-alphas}
\end{align}
and $\mathcal{S}_{1,\cdots,11}$ are given in Appendix~\ref{app:calS}.

For the absence of gradient instabilities, it is necessary and sufficient to impose
\begin{align}
&\alpha_1 >0\,, \
  \alpha_2 >0\,, \
  \alpha_3 >0\,, \
  \alpha_4 >0\,, \
  \alpha_5 >0\,, \nonumber\\
 & \mbox{ for }
 {}^{\forall}p_{\perp} \mbox{ and } {}^{\forall}p_x \mbox{ such that } p^2\ne 0\,,
  \label{eqn:stability1pre}
\end{align}
and
\begin{align} 
 & \alpha_1^2 - \alpha_2 > 0\,,\quad
  \alpha_3^2 - \alpha_4 > 0\,,\nonumber\\
 & -\alpha_5^2 + 2(3\alpha_4-2\alpha_3^2)\alpha_3\alpha_5 + (3\alpha_3^2-4\alpha_4)\alpha_4^2 > 0\,,\nonumber\\
 & \mbox{ for }
 {}^{\forall}p_{\perp} \mbox{ and } {}^{\forall}p_x \mbox{ such that } p^2\ne 0\,.
 \label{eqn:stability2}
\end{align}
Here, we supposed that the no-ghost condition (\ref{eqn:noghost}) is satisfied. The condition (\ref{eqn:stability1pre}) is equivalent to
\begin{equation}
 \mathcal{S}_a >0\,, \quad (a=1,\cdots,11)\,. \label{eqn:stability1}
\end{equation}

\section{Long wavelength perturbations}
\label{sec:longwavelengthperturbations}

In the previous section we studied the stability of the de Sitter solution with a homogeneous magnetic field in the UV limit, i.e. the limit where the wavelengths of perturbations are sufficiently shorter than the size of the de Sitter horizon. In this section we study the stability of the solution in the opposite limit, namely the limit where the wavelengths of perturbations are sufficiently longer than the size of the horizon.

As argued in \cite{Gumrukcuoglu:2016jbh}, instabilities for modes with $p^2=\mathcal{O}(H_0^2)$ can be as harmless as the standard Jeans instability. It is nonetheless necessary to impose the stability for those modes with $p^2\ll H_0^2$ if one wants the background under investigation to realize dynamically as an attractor of the system. In general relativity, the standard Jeans instability prevents the matter-dominated FLRW background from being an attractor of the system, but this is not a problem since it is not the matter-dominated epoch but the inflationary epoch that sets the initial condition of our local patch of the universe. On the other hand, for the system under investigation in the present paper, we would like the homogeneous magnetic field to realize dynamically and to be sustained during a de Sitter phase representing the inflationary epoch. For this reason, we require the solution to be an attractor of the system. Namely, we require the solution to be stable against perturbations with $p^2\ll H_0^2$.

For this purpose we send $p_{\perp}/H_0$ and $p_x/H_0$ to zero after obtaining equations of motion for properly normalized dynamical variables. Since the vector harmonics $Y^{\rm odd/even}_{n, p}$ ($n = cc, cs, sc, ss$; $p, q = 2, 3$) vanish in the limit and $Y^{\rm odd/even}_{n, p}/k_{\perp}$ remain finite, we introduce $\tilde{h}^{\rm odd}_{1,n} \equiv k_{\perp}h^{\rm odd}_{1,n}$, $\tilde{A}^{\rm odd}_n \equiv k_{\perp}A^{\rm odd}_n$ and $\tilde{V}^{\rm odd/even}_n \equiv k_{\perp}V^{\rm odd/even}_n$. The stability of the background solution against anisotropic perturbations with long wavelengths can then be analyzed by studying the equations of motion for the dynamical variables ($\tilde{h}^{\rm odd}_{1,n}$, $A_{1,n}$, $\pi_n$, $\psi_n$, $\tilde{A}^{\rm odd}_n$) in the limit where $p_{\perp}/H_0$ and $p_x/H_0$ are sent to zero, after eliminating non-dynamical variables. For simplicity we set $N(t)=1$ in the rest of this section.

The equation of motion for $\tilde{h}^{\rm odd}_{1,n}$ in the long wavelength limit does not involve other variables and is
\begin{equation}
 \ddot{\tilde{h}}^{\rm odd}_{1,n} + 3H_0\dot{\tilde{h}}^{\rm odd}_{1,n} = 0\,,
\end{equation}
where an over-dot represents derivative with respect to the time variable $t$. This gives the solution
\begin{equation}
 \tilde{h}^{\rm odd}_{1,n} = C_1 + C_2 e^{-3H_0 t}\,,
\end{equation}
where $C_1$ and $C_2$ are constants. The first solution, which is constant in space and time, is actually a pure gauge. Therefore, $\tilde{h}^{\rm odd}_{1,n}$ is stable. 

The equation of motion for $A_{1,n}$ in the long wavelength limit also does not involve other variables and is
\begin{equation}
 \ddot{A}_{1,n} + 7H_0\dot{A}_{1,n} + 6H_0^2A_{1,n} = 0\,,
\end{equation}
which gives the solution
\begin{equation}
 A_{1,n} = C_3e^{-H_0t} + C_4e^{-6H_0t}\,,
\end{equation}
where $C_3$ and $C_4$ are constants. Therefore, $A_{1,n}$ is stable. 

The equations of motion for $\pi_n$ and $\psi_n$ in the long wavelength limit are coupled but can be solved easily by setting
\begin{equation}
 \pi_n = \pi_{n}^{0} e^{\Gamma H_0 t}\,, \quad
   \psi_n = \psi_{n}^{0} e^{\Gamma H_0 t}\,, \label{eqn:pi-psi-homogeneous-anisotropic}
\end{equation}
where $\pi_{n}^{0}$ and $\psi_{n}^{0}$ are constants. The general solution is of course a linear combination of solutions of this form for all allowed values of $\Gamma$. By substituting (\ref{eqn:pi-psi-homogeneous-anisotropic}) to the equations of motion for $\pi_n$ and $\psi_n$, one obtains
\begin{equation}
 \Gamma (\Gamma+3)\left(\Gamma^2 + 3\Gamma + \frac{\mathcal{A}}{\mathcal{N}}\right) = 0\,,
\end{equation}
where $\mathcal{A}$ and $\mathcal{N}$ are the same as those defined in (\ref{eqn:def-calN-calA}) and thus in particular $\mathcal{N}=\mathcal{N}_5$ (see (\ref{eqn:def-calN1to5})). For $\Gamma=0$, $\pi_{n}^{0}=0$ and $\psi_n$ is a pure gauge. Therefore, $\pi_n$ and $\psi_n$ are also stable, provided that the attractor condition (\ref{eqn:attractor}) and the no-ghost condition (\ref{eqn:noghost}) are satisfied.

The equation of motion for $\tilde{A}^{\rm odd}_n$ in the long wavelength limit does not involve other variables and is
\begin{align}
& \ddot{\tilde{A}}^{\rm odd}_n + 5H_0\dot{\tilde{A}}^{\rm odd}_n
 + \mathcal{B}H_0^2 \tilde{A}^{\rm odd}_n = 0\,, \nonumber\\
& \mathcal{B} = \frac{\mathcal{A}_1 k_x^4 + 2(\mathcal{A}_3-\sqrt{\mathcal{A}_1\mathcal{A}_2})k_x^2k_{\perp}^2 + \mathcal{A}_2k_{\perp}^4}{[(\zeta_1-\zeta_3)k_x^2+(3g_hb^2+\zeta_1-\zeta_3)k_{\perp}^2]^2}\,, \label{eqn:eom-tildeAodd-longwavelength}
\end{align}
where $\mathcal{A}_{1,2,3}$ are given in Appendix~\ref{app:calA123}. Therefore, $\tilde{A}^{\rm odd}_n$ is stable if and only if
\begin{equation}
 \mathcal{A}_a > 0\,, \quad (a=1,2,3)\,. \label{eqn:stability-anisotropy}
\end{equation}

\section{Examples}
\label{sec:examples}

In this section we first show a couple of classes of models that always violate either (\ref{eqn:noghost}) or (\ref{eqn:stability1}). After that, we show some concrete examples of stable models with specific choices of parameters, which satisfy not only (\ref{eqn:noghost}) and (\ref{eqn:stability1}) but also (\ref{eqn:stability2}), (\ref{eqn:attractor}) and (\ref{eqn:stability-anisotropy}).

\subsection{Unstable classes of models}

Let us first consider the case without the Horndeski's non-minimal vector coupling ($\xi = 0$). Setting $g_h = 0$, we obtain $\mathcal{S}_2 = 0$. This implies that $\alpha_2 = 0$ and thus one of the five modes has vanishing sound speed.

Let us next consider the case with the Horndeski's non-minimal vector coupling ($\xi \ne 0$) but without Horndeski scalar terms. Setting $g_{3x} = g_{3xx} = g_{4x} = g_{4xx} = g_{5x} = g_{5xx} = 0$, we obtain $\mathcal{S}_9/\mathcal{S}_2 = -(2\mathcal{N}_2-\mathcal{N}_1)^2\mathcal{N}_2$, which is non-positive if we impose the no-ghost condition. Therefore, either $\mathcal{S}_9$ or $\mathcal{S}_2$ is non-positive if the no-ghost condition is imposed.

These two unstable examples imply that the stability requires both the Horndeski's nonminimal vector coupling and Horndeski scalar terms.

\subsection{Stable models}
 
 Our strategy here is to find a set of parameters that satisfies (\ref{eqn:noghost}) and (\ref{eqn:stability1}), and then to check if (\ref{eqn:stability2}), (\ref{eqn:attractor}) and (\ref{eqn:stability-anisotropy}) are also satisfied.

We have already assumed (\ref{eqn:Parity}). For simplicity, in all the examples considered in this subsection we further set
\begin{align}
0 =& g_{3xx} = g_{4xx} = g_{4xxx} = g_{5x} \nonumber\\
 =&  g_{5xx} = g_{5xxx} = l_{yy} = l_{xw} = l_{z}\,. \label{eqn:simplicity}
\end{align}
In each of the following examples we specify ($g_h$, $g_4$, $g_{4x}$, $g_{3x}$, $b$, $l_{xx}$, $l_{ww}$). We then solve the background equations of motion (\ref{eqn:dSnoE}) with respect to ($l$, $l_x$, $l_w$), where
\begin{equation}
 L = l\, M_{\rm Pl}^2H_0^2\,, \
  L_X = l_x\, M_{\rm Pl}^2\,, \ 
  L_W = l_w\, M_{\rm Pl}^2\,, \label{eqn:def-l-lx-lw}
\end{equation}
and it is understood that the left hand sides of (\ref{eqn:def-l-lx-lw}) are evaluated at the de Sitter solution with a homogeneous magnetic field. After that, we confirm that (\ref{eqn:noghost}) and (\ref{eqn:stability1}) as well as (\ref{eqn:stability2}), (\ref{eqn:attractor}) and (\ref{eqn:stability-anisotropy}) are satisfied.

For the set of parameters ($g_h$, $g_4$, $g_{4x}$, $g_{3x}$, $b$, $l_{xx}$, $l_{ww}$; $l$, $l_x$, $l_w$) and the assumptions (\ref{eqn:Parity}) and (\ref{eqn:simplicity}), one can easily reconstruct the Lagrangian assuming a simple ansatz and noting that $X=2H_0^2$ and $W=-b^2H_0^2/2$ on the de Sitter solution parameterized by ($H_0$, $b$). For example, if we assume that $L$ is quadratic in $X$ and $W$ and independent of $Y$ and $Z$ then 
\begin{align}
 \frac{L}{M_{\rm Pl}^2} =&
  \left(l-2 l_x+2 l_{xx}+\frac{1}{2} b^2 l_w+\frac{1}{8} b^4 l_{ww}\right) H_0^2 \nonumber\\
&
  +(l_x-2 l_{xx})X + \left(l_w+\frac{1}{2} b^2 l_{ww}\right)W \nonumber\\
&
  +\frac{1}{2H_0^2}( l_{xx} X^2 + l_{ww} W^2)\,.  
\end{align}
The simplest choice of $G_{3,4,5}$ is
\begin{equation}
 G_3 = g_{3x}\frac{M_{\rm Pl}^2}{H_0^2}X\,,\ 
 G_4 = g_4M_{\rm Pl}^2 + g_{4x}\frac{M_{\rm Pl}^2}{H_0^2}X\,,\ 
 G_5 = 0\,.
\end{equation}
The value of $\xi$ is 
\begin{equation}
 \xi = g_h \frac{H_0^2}{M_{\rm Pl}^2}\,.
\end{equation}

\subsubsection{Example 1} 

Let us consider the following choice of parameters. 
\begin{align}
&
  g_h = -1\,, \
  g_4 = \frac{1}{2}\,, \
  g_{4x} = 0\,,\ 
  g_{3x} = \frac{1}{10}\,, \nonumber\\
&
  b = \frac{1}{10}\,, \
  l_{xx} = -\frac{1}{10}\,, \
  l_{ww} = -1\,,
\end{align}
for which the background equations of motion (\ref{eqn:dSnoE}) give
\begin{equation}
 l = -\frac{79}{25}\,,\ l_x = -\frac{31}{50}\,,\ l_w = 20\,. \label{eqn:l-lx-lw-ex1}
\end{equation}

In this case we have
\begin{align}
& \mathcal{N}_1 = \frac{12}{25}\,, \
  \mathcal{N}_2 = \frac{1}{2}\,, \
  \mathcal{N}_3 = 12\,, \nonumber\\
& \mathcal{N}_4 = 12\,, \
  \mathcal{N}_5 = \frac{1408}{15625}\,,
\end{align}
and
\begin{align}
 &
  \mathcal{S}_1 = \frac{1}{2}\,,\
  \mathcal{S}_2 = \frac{1}{50}\,, \ 
  \mathcal{S}_3 = \frac{59456}{9765625}\,, \nonumber\\
 &
  \mathcal{S}_4 = \frac{35469972}{6103515625}\,, \ 
  \mathcal{S}_5 = \frac{1696}{390625}\,, \
  \mathcal{S}_6 = \frac{7940769}{1953125000}\,, \nonumber\\
 &
  \mathcal{S}_7 = \frac{2060321}{488281250}+\sqrt{\mathcal{S}_5\mathcal{S}_6}\,,\
  \mathcal{S}_8 = \frac{1201}{100}\,, \
  \mathcal{S}_9 = \frac{224}{15625}\,, \nonumber\\
 &
  \mathcal{S}_{10} = \frac{29}{2500}\,, \
 \mathcal{S}_{11} = \frac{823}{62500}+\sqrt{\mathcal{S}_9\mathcal{S}_{10}}\,.
  \label{eqn:calS-example1}
\end{align}
Thus (\ref{eqn:noghost}) and (\ref{eqn:stability1}) are satisfied.

The conditions (\ref{eqn:attractor}) and (\ref{eqn:stability-anisotropy}) are also satisfied as
\begin{align}
 &
  \mathcal{A} = \frac{5979}{390625}\,,\
  \mathcal{A}_1 = \frac{256}{625}\,,\
  \mathcal{A}_2 = \frac{87451}{250000}\,,\nonumber\\
&
  \mathcal{A}_3 = \frac{47493}{125000}+\sqrt{\mathcal{A}_1\mathcal{A}_2}\,. 
\end{align}

After substituting (\ref{eqn:calS-example1}) into (\ref{eqn:def-alphas}) to obtain explicit expressions of $\alpha_{1,\cdots,5}$, it is straightforward to compute the three expressions on the left hand sides of the inequalities in (\ref{eqn:stability2}) and to show that they all have the form $A/B$, where $A$ and $B$ are polynomials of $p_{\perp}^2$ and $p_x^2$ with positive coefficients. Therefore, (\ref{eqn:stability2}) is also satisfied.

\subsubsection{Example 2}

Let us consider the following choice of parameters. 
\begin{align}
&
  g_h = -1\,, \
  g_4 = \frac{1}{2}\,, \
  g_{4x} = 0\,,\ 
  g_{3x} = \frac{1}{10}\,, \nonumber\\
&
  b = \frac{1}{10}\,, \
  l_{xx} = -\frac{1}{10}\,, \
  l_{ww} = 0\,,
\end{align}
for which the background equations of motion (\ref{eqn:dSnoE}) give the same values of $l$, $l_x$ and $l_w$ as shown in (\ref{eqn:l-lx-lw-ex1}).

In this case we have
\begin{align}
& \mathcal{N}_1 = \frac{12}{25}\,, \
  \mathcal{N}_2 = \frac{1}{2}\,, \
  \mathcal{N}_3 = 12\,, \nonumber\\
& \mathcal{N}_4 = 12\,, \
  \mathcal{N}_5 = \frac{1408}{15625}\,,
\end{align}
and
\begin{align} \label{eqn:calS-example3}
&
  \mathcal{S}_1 = \frac{1}{2}\,,\
  \mathcal{S}_2 = \frac{1}{50}\,, \ 
  \mathcal{S}_3 = \frac{59456}{9765625}\,, \nonumber\\
&
  \mathcal{S}_4 = \frac{1418292}{244140625}\,, \ 
  \mathcal{S}_5 = \frac{1696}{390625}\,, \
  \mathcal{S}_6 = \frac{79353}{19531250}\,, \nonumber\\
&
  \mathcal{S}_7 = \frac{16477}{3906250}+\sqrt{\mathcal{S}_5\mathcal{S}_6}\,,\
  \mathcal{S}_8 = 12\,, \
  \mathcal{S}_9 = \frac{224}{15625}\,, \nonumber\\
&
  \mathcal{S}_{10} = \frac{29}{2500}\,, \
  \mathcal{S}_{11} = \frac{823}{62500}+\sqrt{\mathcal{S}_9\mathcal{S}_{10}}\,. 
\end{align}
Thus (\ref{eqn:noghost}) and (\ref{eqn:stability1}) are satisfied.

The conditions (\ref{eqn:attractor}) and (\ref{eqn:stability-anisotropy}) are also satisfied as
\begin{align}
&
  \mathcal{A} = \frac{239}{15625}\,,\
  \mathcal{A}_1 = \frac{256}{625}\,,\
  \mathcal{A}_2 = \frac{1749}{5000}\,,\nonumber\\
&
  \mathcal{A}_3 = \frac{5699}{15000}+\sqrt{\mathcal{A}_1\mathcal{A}_2}\,. 
\end{align}

After substituting (\ref{eqn:calS-example2}) into (\ref{eqn:def-alphas}) to obtain explicit expressions of $\alpha_{1,\cdots,5}$, it is straightforward to compute the three expressions on the left hand sides of the inequalities in (\ref{eqn:stability2}) and to show that they all have the form $A/B$, where $A$ and $B$ are polynomials of $p_{\perp}^2$ and $p_x^2$ with positive coefficients. Therefore, (\ref{eqn:stability2}) is also satisfied.

\subsubsection{Example 3}

Let us consider the following choice of parameters. 
\begin{align}
&
  g_h = -\frac{1}{20}\,, \
  g_4 = \frac{1}{2}\,, \
  g_{4x} = 0\,,\ 
  g_{3x} = \frac{1}{10}\,, \nonumber\\
&
  b = \frac{1}{10}\,, \
  l_{xx} = -\frac{1}{10}\,, \
  l_{ww} = 0\,,
\end{align}
for which the background equations of motion (\ref{eqn:dSnoE}) give
\begin{equation}
 l = -\frac{376}{125}\,,\ l_x = -\frac{601}{1000}\,,\ l_w = 1\,.
\end{equation}

In this case we have
\begin{align}
& \mathcal{N}_1 = \frac{499}{1000}\,, \
  \mathcal{N}_2 = \frac{1}{2}\,, \
  \mathcal{N}_3 = \frac{3}{5}\,, \nonumber\\
& \mathcal{N}_4 = \frac{3}{5}\,, \
  \mathcal{N}_5 = \frac{108990799}{1000000000}\,,
\end{align} 
and
\begin{align} \label{eqn:calS-example2}
&
  \mathcal{S}_1 = \frac{1}{2}\,,\
  \mathcal{S}_2 = \frac{1}{1000}\,, \ 
  \mathcal{S}_3 = \frac{371370384707}{1000000000000000}\,, \nonumber\\
&
  \mathcal{S}_4 = \frac{92615485451247}{250000000000000000}\,, \ 
  \mathcal{S}_5 = \frac{50621001581}{200000000000000}\,, \nonumber\\
&
  \mathcal{S}_6 = \frac{252272242903}{1000000000000000}\,, \nonumber\\
&
  \mathcal{S}_7 = \frac{50538124821}{200000000000000}+\sqrt{\mathcal{S}_5\mathcal{S}_6}\,,\
  \mathcal{S}_8 = \frac{3}{5}\,, \nonumber\\
&
  \mathcal{S}_9 = \frac{29949199}{2000000000}\,, \
    \mathcal{S}_{10} = \frac{59301}{4000000}\,, \nonumber\\
&
  \mathcal{S}_{11} = \frac{59601699}{4000000000}+\sqrt{\mathcal{S}_9\mathcal{S}_{10}}\,.
\end{align}
Thus (\ref{eqn:noghost}) and (\ref{eqn:stability1}) are satisfied.

The conditions (\ref{eqn:attractor}) and (\ref{eqn:stability-anisotropy}) are also satisfied as
\begin{align}
&
  \mathcal{A} = \frac{304101}{500000000}\,,\
  \mathcal{A}_1 = \frac{90601}{250000}\,,\
  \mathcal{A}_2 = \frac{718809}{2000000}\,,\nonumber\\
&
  \mathcal{A}_3 = \frac{90045653}{249500000}+\sqrt{\mathcal{A}_1\mathcal{A}_2}\,. 
\end{align}

After substituting (\ref{eqn:calS-example3}) into (\ref{eqn:def-alphas}) to obtain explicit expressions of $\alpha_{1,\cdots,5}$, it is straightforward to compute the three expressions on the left hand sides of the inequalities in (\ref{eqn:stability2}) and to show that they all have the form $A/B$, where $A$ and $B$ are polynomials of $p_{\perp}^2$ and $p_x^2$ with positive coefficients. Therefore, (\ref{eqn:stability2}) is also satisfied.

\subsubsection{Example 4}

Let us consider the following choice of parameters. 
\begin{align}
&
  g_h = -\frac{1}{20}\,, \
  g_4 = \frac{1}{2}\,, \
  g_{4x} = -\frac{1}{10}\,,\ 
  g_{3x} = 0\,, \nonumber\\
&
  b = \frac{1}{10}\,, \
  l_{xx} = -\frac{1}{10}\,, \
  l_{ww} = 0\,,
\end{align}
for which the background equations of motion (\ref{eqn:dSnoE}) give
\begin{equation}
 l = -\frac{676}{125}\,,\ l_x = \frac{599}{1000}\,,\ l_w = 1\,.
\end{equation}

In this case we have
\begin{align}
& \mathcal{N}_1 = \frac{899}{1000}\,, \
  \mathcal{N}_2 = \frac{9}{10}\,, \
  \mathcal{N}_3 = \frac{3}{5}\,, \nonumber\\
& \mathcal{N}_4 = \frac{3}{5}\,, \
  \mathcal{N}_5 = \frac{108704599}{1000000000}\,,
\end{align} 
and
\begin{align} \label{eqn:calS-example4}
&
  \mathcal{S}_1 = \frac{1}{2}\,,\
  \mathcal{S}_2 = \frac{1}{1000}\,, \ 
  \mathcal{S}_3 = \frac{9143271446151}{5000000000000000}\,, \nonumber\\
&
  \mathcal{S}_4 = \frac{456009833453647}{250000000000000000}\,, \ 
  \mathcal{S}_5 = \frac{9305798288929}{5000000000000000}\,, \nonumber\\
&
  \mathcal{S}_6 = \frac{9284212649127}{5000000000000000}\,, \nonumber\\
&
  \mathcal{S}_7 = \frac{9295017384133}{5000000000000000}+\sqrt{\mathcal{S}_5\mathcal{S}_6}\,,\
  \mathcal{S}_8 = \frac{3}{5}\,,  \nonumber\\
&
  \mathcal{S}_9 = \frac{207508199}{2000000000}\,, \
    \mathcal{S}_{10} = \frac{10351541}{100000000}\,, \nonumber\\
&
  \mathcal{S}_{11} = \frac{414541899}{4000000000}+\sqrt{\mathcal{S}_9\mathcal{S}_{10}}\,.
\end{align}
Thus (\ref{eqn:noghost}) and (\ref{eqn:stability1}) are satisfied.

The conditions (\ref{eqn:attractor}) and (\ref{eqn:stability-anisotropy}) are also satisfied as
\begin{align}
&
  \mathcal{A} = \frac{9185101}{500000000}\,,\
  \mathcal{A}_1 = \frac{1692601}{250000}\,,\
  \mathcal{A}_2 = \frac{13530009}{2000000}\,,\nonumber\\
&
  \mathcal{A}_3 = \frac{3042086553}{449500000}+\sqrt{\mathcal{A}_1\mathcal{A}_2}\,. 
\end{align}

After substituting (\ref{eqn:calS-example4}) into (\ref{eqn:def-alphas}) to obtain explicit expressions of $\alpha_{1,\cdots,5}$, it is straightforward to compute the three expressions on the left hand sides of the inequalities in (\ref{eqn:stability2}) and to show that they all have the form $A/B$, where $A$ and $B$ are polynomials of $p_{\perp}^2$ and $p_x^2$ with positive coefficients. Therefore, (\ref{eqn:stability2}) is also satisfied.

\section{Summary and discussion}
\label{sec:summary}

We have presented a detailed stability analysis for the de Sitter solution with a homogeneous magnetic field that was recently found in \cite{Mukohyama:2016npi} in the context of a $U(1)$ gauge theory nonminimally coupled to scalar-tensor gravity. The magnetic field is ``stealth'' in the sense that the corresponding stress-energy tensor is of the form of an effective cosmological constant and thus is isotropic despite the fact that the magnetic field has a preferred spatial direction. We have studied the stability of the solution against linear perturbations in the subhorizon and superhorizon limits and have shown some explicit examples that satisfy all stability conditions. Stable models include both Horndeski's nonminimal vector coupling and Horndeski scalar terms.

The stable de Sitter solution with a homogeneous magnetic field opens up a new possibility for inflationary magnetogenesis, in which magnetic fields in the Universe at all scales may originate from a classical, homogeneous magnetic field sustained during inflation. Towards such a new scenario of inflationary magnetogenesis, an important step forward is to show a graceful exit from the de Sitter solution with constant values of ($X$, $W$, $Y$, $Z$, $\chi$). In particular, the scalar field $\phi$ should be stabilized around a local minimum of its potential with a sufficiently large mass at late time after inflation in order to recover the standard Einstein-Maxwell theory. (To be more precise, what is recovered is the Einstein-Maxwell theory with the Horndeski's nonminimal vector coupling.) Also, as already pointed out in \cite{Mukohyama:2016npi}, a field other than $\phi$ needs to be introduced as the main source of curvature perturbation, i.e. either an inflaton or a curvaton, in order to avoid too large statistical anisotropies and non-Gaussianities. One eventually needs to establish the stability of the whole system all the way from the inflationary epoch to the standard expansion history at late time through the graceful exit. The present paper has established the stability of the inflationary epoch and thus can be considered as the first step towards the new inflationary magnetogenesis scenario based on a classical, homogeneous magnetic field sustained during inflation.

The recent multi-messenger detection of binary neutron stars put a strong constraint on the deviation of the propagation speed of gravitational waves from that of light at the present~\cite{TheLIGOScientific:2017qsa,Monitor:2017mdv}. If the Horndeski scalar-tensor theory is used as a model of the late-time acceleration of the universe then the functions $G_4$ and $G_5$ are strongly constrained. On the other hand, the propagation speed of gravitational waves in the early universe such as the inflationary epoch is not constrained at all by such observations. Therefore, the model considered in \cite{Mukohyama:2016npi} and in the present paper is consistent with the multi-messenger detection of binary neutron stars as far as the scalar field is stabilized around a local minimum of its potential with a sufficiently large mass at late time after inflation, as required anyway for the recovery of the Einstein-Maxwell theory (with the Horndeski's nonminimal vector coupling).

After inflation and the stabilization of $\phi$ around a local minimum of its potential, the homogeneous magnetic field is no longer stealth and thus the background spacetime is then described by a Bianchi I geometry (instead of FLRW one). It is therefore important to investigate phenomenology of the Bianchi I universe with a homogeneous magnetic field and the standard content of the universe (radiation, matter and the cosmological constant) in the context of the Einstein-Maxwell theory with the Horndeski's nonminimal vector coupling~\cite{nextproject}.

\begin{acknowledgments}
 The author thanks Axel Brandenburg, Ruth Durrer, Yiwen Huang, Tina Kahniashvili and Sayan Mandal for helpful discussions. A part of the work was completed during the author's visits to Simon Fraser University, University of Alberta and University of Victoria. The author is therefore grateful to Andrei Frolov, Valeri Frolov, Werner Israel, Maxim Pospelov and Andrei Zelnikov for their warm hospitality. The work was supported in part by Japan Society for the Promotion of Science (JSPS)  Grants-in-Aid for Scientific Research (KAKENHI) No. 17H02890, No. 17H06359, and by World Premier International Research Center Initiative (WPI), MEXT, Japan.
\end{acknowledgments}

\appendix

\section{Explicit expressions of sub-matrices}
\label{app:submatrices}

The sub-matrices in (\ref{eqn:submatrices}) are 
 \begin{align}
{\bf K}_{\rm odd} =&
  \left(\begin{array}{cc}
   \zeta_6 & 0 \\
	 0 & \zeta_1
	  \end{array}
			  \right)\,, \\
{\bf K}_{\rm even} =&
  \left(\begin{array}{ccc}
   \zeta_7 & 0 & 0 \\
	 0 & \zeta_2 & \zeta_3\\
	 0 & \zeta_3 &  0
	  \end{array}
  \right)\,, \\
{\bf M}_{\rm even} =& H_0
  \left(\begin{array}{ccc}
   0 & 0 & 0 \\
	 0 & 0 & -b^2g_h\\
	 0 & b^2g_h &  0
	  \end{array}
 \right)\,, \\
{\bf M}_{\rm mix} =& p_{\perp}
  \left(\begin{array}{ccc}
   0 & w_1 & 2bg_h\\
	 -2bg_h & 0 & 0
	 \end{array}
  \right)\,, \\
{\bf V}_{\rm odd}  =&
  p_{\perp}^2
  \left(\begin{array}{cc}
   \zeta_5 & 0\\
	 0 & g_4
	  \end{array}
					 \right)
  +  p_{x}^2
  \left(\begin{array}{cc}
   -12g_h & 0\\
	 0 & 0
	\end{array}
  \right)\nonumber\\
 &
  +H_0^2
  \left(\begin{array}{cc}
   -4\zeta_6& 0\\
	 0 & 0
	\end{array}
				  \right)\,, \\
{\bf V}_{\rm even} =& 
 p_{\perp}^2
 \left(\begin{array}{ccc}
   -12g_h & 0 & 0\\
	 0 & w_2 & w_3\\
	 0 & w_3 & 0
       \end{array}
				 \right)
 + p_{x}^2
 \left(\begin{array}{ccc}
  0 & 0 & 0\\
	 0 & w_4 & 0\\
	 0 & 0 & 0
       \end{array}
	 \right)\nonumber\\
& +  H_0^2
 \left(\begin{array}{ccc}
  -4\zeta_7 & 0 & 0\\
	 0 & w_5 & 3b^2g_h\\
	 0 & 3b^2g_h & 0
       \end{array}
			  \right)\,, \\
{\bf V}_{\rm mix} =& 
  p_{\perp}H_0
 \left(\begin{array}{ccc}
  0 & w_6 & 2bg_h\\
	8bg_h & 0 & 0
       \end{array}
			  \right)\,, \\
{\bf A}_{\rm odd}  =&
  p_{x}H_0
  \left(\begin{array}{cc}
   0 & -4b^2g_h
	\end{array}
				  \right)\,,\\
{\bf A}_{\rm even} =&
 \frac{p_{\perp}^3}{H_0}
 \left(\begin{array}{ccc}
  0 & 0 & 0\\
	0 & 0 & 0\\
	   0 & 0 & 0\\
	0 & 8b^2g_h-2\zeta_3 & \zeta_1
       \end{array}
						   \right)\nonumber\\
 & 
 + \frac{p_{\perp}p_{x}^2}{H_0}
 \left(\begin{array}{ccc}
  0 & 0 & 0\\
	0 & 0 & 0\\
	   0 & 0 & 0\\
	0 & -2\zeta_3 & 0
       \end{array}
	 \right)\nonumber\\
& + p_{\perp}H_0
 \left(\begin{array}{ccc}
  0 & b^2\left(2g_h+\frac{1}{2}\zeta_6\right) & -2b^2g_h\\
	0 & 0 & 0 \\
	0 & 0 & 0 \\
	0 & -b\zeta_4 & 0
       \end{array}
	 \right)\nonumber\\
& + p_{x}H_0
 \left(\begin{array}{ccc}
	0 & 0 & 0 \\
	-\zeta_7 & 0 & 0 \\
	0 & -4b^2g_h & 0 \\
	0 & 0 & 0
       \end{array}
	 \right)\,, \\
{\bf A}_{\rm mix I} =& 
  p_{\perp}p_{x}
 \left(\begin{array}{ccc}
  4bg_h & 0 & 0
       \end{array}
  \right)\,, \\
{\bf A}_{\rm mix II} =&
  p_{\perp}^2
 \left(\begin{array}{cc}
  0 & 0\\
	0 & 0\\
	0 & 0\\
	\zeta_4 & 0
       \end{array}
  \right)
 +  p_{\perp}p_{x}
 \left(\begin{array}{cc}
  0 & 0\\
	0 & 0\\
	4bg_h & 0\\
	0 & 0
       \end{array}
  \right) \nonumber\\
& +  p_{x}^2
 \left(\begin{array}{cc}
  4bg_h & 0\\
	0 & 0\\
	0 & 0\\
	0 & 0
       \end{array}
  \right)
 +  H_0^2
 \left(\begin{array}{cc}
  b\zeta_6 & 0\\
	0 & 0\\
	0 & 0\\
	0 & 0
       \end{array}
  \right)\,,\\
{\bf B}_{\rm odd}  =&
  p_{x}
  \left(\begin{array}{cc}
   0 & -\zeta_1
	\end{array}
				  \right)\,,\\
{\bf B}_{\rm even} =&
 p_{\perp}
 \left(\begin{array}{ccc}
  0 & -8b^2g_h+2\zeta_3 & -\zeta_1\\
	0 & 0 & 0\\
	   0 & 0 & 0\\
	0 & w_7 & 2(\zeta_1-\zeta_3)
       \end{array}
						   \right)\nonumber\\
 & 
 + p_{x}
 \left(\begin{array}{ccc}
  0 & 0 & 0\\
	\zeta_7 & 0 & 0\\
	   0 & -2\zeta_3 & 0\\
	0 & 0 & 0
       \end{array}
	 \right)\,, \\
{\bf B}_{\rm mix} =&
  H_0
 \left(\begin{array}{cc}
  -b\zeta_6 & 0\\
	0 & 0\\
	0 & 0\\
	0 & 0
       \end{array}
		      \right)\,, \\
{\bf C}_{\rm odd}   =&
  p_{\perp}^2
  \left(\begin{array}{c}
   -2b^2g_h+\zeta_1
	\end{array}
					 \right)
  +  p_{x}^2
  \left(\begin{array}{c}
   \zeta_1
	\end{array}
  \right)
  +H_0^2
  \left(\begin{array}{c}
   b^2\zeta_6
	\end{array}
				  \right)\,,\\
{\bf C}_{\rm even} =&
 p_{\perp}^2
 \left(\begin{array}{cccc}
   0 & 0 & 0 & w_8\\
	0 & \zeta_6 & 0 & 0\\
	0 & 0 & \zeta_1 & 0\\
	w_8 & 0 & 0 & w_9
       \end{array}
						   \right)\nonumber\\
  & 
 + p_{\perp}p_{x}
 \left(\begin{array}{cccc}
  0 & 0 & \zeta_1 & 0\\
	0 & 0 & 0 & 0\\
	\zeta_1 & 0 & 0 & 4(\zeta_3-\zeta_1)\\
	0 & 0 & 4(\zeta_3-\zeta_1) & 0
       \end{array}
	 \right)\nonumber\\
 & + p_{x}^2
 \left(\begin{array}{cccc}
  \zeta_1 & 0 & 0 & 0\\
	0 & \zeta_7 & 0 & 0\\
	0 & 0 & 0 & 0\\
	0 & 0 & 0 & 0
       \end{array}
	 \right)
+  H_0^2
 \left(\begin{array}{cccc}
  b^2\zeta_6 & 0 & 0 & 0\\
	0 & 0 & 0 & 0\\
	0 & 0 & 0 & 0\\
	0 & 0 & 0 & 0
       \end{array}
			  \right)\,, \\
{\bf C}_{\rm mix} =&
  p_{\perp}H_0
 \left(\begin{array}{cccc}
  0 & b\zeta_6 & 0 & 0
       \end{array}
			  \right)\,, 
\end{align}
where
 \begin{align}
 w_1 =& -b\left(2g_h+\frac{1}{4}\zeta_6\right)+\frac{1}{4}\zeta_4\,,\nonumber\\
 w_2 =& -b^2\left(7g_h+\frac{1}{4}\zeta_6\right)+\zeta_1+\zeta_3-g_4\,,\nonumber\\
 w_3 =& b^2g_h-\frac{1}{2}\zeta_1+\frac{1}{2}g_4\,,\nonumber\\
 w_4 =& -4b^2g_h +\zeta_1+\zeta_3-g_4\,,\nonumber\\
 w_5 =& (-6g_h+\zeta_5)b^2+\frac{3}{2}b\zeta_4\,,\nonumber\\
 w_6 =& -b(2g_h+\zeta_5+\zeta_6) - \frac{1}{2}\zeta_4\,,\nonumber\\
 w_7 =& 16b^2g_h-2(\zeta_2+3\zeta_3)\,,\nonumber\\
 w_8 =& 12b^2g_h+4(\zeta_1-\zeta_3)\,,\nonumber\\
 w_9 =& -56b^2g_h+4(-3\zeta_1+\zeta_2+6\zeta_3)\,.
\end{align}

\section{Coefficients of matrix components}
\label{app:coefficients}

The coefficients $c_1, \cdots, c_{17}$ in (\ref{eqn:tildeK-tildeM-tildeV}) are as follows. 
 \begin{align}
 c_1 =& \zeta_1(\zeta_1\zeta_2+3\zeta_3^2)- 2b^2g_h(\zeta_1\zeta_2+8\zeta_1\zeta_3+2\zeta_3^2)\nonumber\\
  & +b^4g_h^2(16\zeta_1+\zeta_2+24\zeta_3)-32b^6g_h^3\,,\\
 c_2 =& \zeta_1(\zeta_1\zeta_2+3\zeta_3^2) - b^2g_h(\zeta_1\zeta_2+16\zeta_1\zeta_3+\zeta_3^2)\nonumber\\
 & +2b^4g_h^2\zeta_3\left(8-\frac{\zeta_3}{\zeta_1}\right)\,,\\
 c_3 =& \frac{1}{4}\zeta_1(\zeta_4-b\zeta_6)+\frac{1}{4}bg_h[-8(\zeta_1-\zeta_3)-b\zeta_4+b^2\zeta_6]\nonumber\\
 & -6b^3g_h^2\,,\\
 c_4 =& \frac{1}{4}\zeta_1(\zeta_4-b\zeta_6)-2bg_h(\zeta_1-\zeta_3)- 4b^3g_h^2\left(4-\frac{\zeta_3}{\zeta_1}\right)\,,\\
 c_5 =& \zeta_1\left[(\zeta_1\zeta_2+3\zeta_3^2)-\frac{b}{8}(\zeta_4-b\zeta_6)(4\zeta_1-\zeta_3)\right]\nonumber\\
 & +b^2g_h\zeta_1(4\zeta_1-\zeta_2-21\zeta_3)+32b^4g_h^2\zeta_1\,,\\
 c_6 =& \zeta_1(\zeta_1\zeta_2+3\zeta_3^2)-2b^2g_h\zeta_3(8\zeta_1-\zeta_3)\,,\\
 c_7 =& -\frac{1}{4}\zeta_1^2\zeta_6b^2 -(\zeta_1-\zeta_3)^2g_4+(\zeta_1-\zeta_3)\zeta_1^2\nonumber\\
 & 
  +\left[\frac{b^2}{2}\zeta_1\zeta_6+6(\zeta_3-\zeta_1)g_4-\zeta_1(\zeta_1-4\zeta_3)\right]b^2g_h \nonumber\\
 & -\left[\frac{1}{4}\zeta_6b^2+3(3\zeta_1+3g_4+\zeta_3)\right]b^4g_h^2+9b^6g_h^3\,,\\
 c_8 =& -\frac{b^2}{2}\zeta_1^2\zeta_6-3(\zeta_1-\zeta_3)^2g_4+3(\zeta_1-\zeta_3)\zeta_1^2 \nonumber\\
  &
  + \left[\frac{b^2}{2}\zeta_1\zeta_6
     +12(\zeta_3-\zeta_1)g_4+6\zeta_1(2\zeta_3-\zeta_1)\right]b^2g_h \nonumber\\
 & 
  -(17\zeta_1+7\zeta_3+9g_4)b^4g_h^2 + 12b^6g_h^3\,,\\
 c_9 =& -\frac{b^2}{4}\zeta_1^2\zeta_6-3(\zeta_1-\zeta_3)^2g_4
  +3(\zeta_1-\zeta_3)\zeta_1^2 \nonumber\\
 &
  + 3\left[2(\zeta_3-\zeta_1) g_4+\zeta_1(4\zeta_3-3\zeta_1)\right]b^2g_h
  \nonumber\\
 & 
-4(2\zeta_1+\zeta_3) b^4g_h^2\,,\\
 c_{10} =&
  -(\zeta_1-\zeta_3)^2g_4 +\zeta_1(\zeta_1-\zeta_3)(\zeta_1-4b^2g_h)\,,\\
 c_{11} =&
  \frac{1}{4}\zeta_1^{2}g_4(\zeta_1\zeta_2+3\zeta_3^2)^2 \nonumber\\
 &
  -\zeta_1(\zeta_1\zeta_2+3\zeta_3^2)(\zeta_1\zeta_2+8\zeta_1\zeta_3+2\zeta_3^2) g_4b^2g_h \nonumber\\
 &
  + \left[ -\frac{1}{16}\zeta_1^2\zeta_6(4\zeta_1-\zeta_3)^2b^2 + (-4\zeta_1^4+10\zeta_1^3\zeta_2\right.\nonumber\\
 & \left.+10\zeta_1^3\zeta_3+\frac{5}{4}\zeta_1^2\zeta_2^2+\frac{51}{2}\zeta_1^2\zeta_3\zeta_2+\frac{343}{4}\zeta_1^2\zeta_3^2\right.\nonumber\\
& \left.
	   +\frac{9}{2}\zeta_1\zeta_3^2\zeta_2+63\zeta_1\zeta_3^3+3\zeta_3^4)g_4+\frac{1}{4}\zeta_1^2(4\zeta_1-\zeta_3) \right.\nonumber\\
 & \left. \times(4\zeta_1^2-2\zeta_1\zeta_2-5\zeta_1\zeta_3-5\zeta_3^2)
 \right]b^4g_h^2
\nonumber\\
 &
  + \left[ \frac{1}{4}\zeta_1\zeta_6(4\zeta_1-\zeta_3)^2b^2
     + (-16\zeta_1^3-30\zeta_1^2\zeta_2 \right. \nonumber\\
 & \left.
      -144\zeta_1^2\zeta_3-\frac{41}{2}\zeta_1\zeta_3\zeta_2-219\zeta_1\zeta_3^2-\frac{1}{2}\zeta_1\zeta_2^2 \right.\nonumber\\
 & \left. -\zeta_3^2\zeta_2-30\zeta_3^3)g_4-\frac{1}{4}\zeta_1(4\zeta_1-\zeta_3)(12\zeta_1^2\right. \nonumber\\
 & \left. -10\zeta_1\zeta_2-59\zeta_1\zeta_3-20\zeta_3^2)\right] b^6g_h^3
\nonumber\\
 &
  + \left[ -\frac{1}{4}\zeta_6(4\zeta_1-\zeta_3)^2b^2
     + (112\zeta_1^2+20\zeta_1\zeta_2\right.\nonumber\\
 & \left. +296\zeta_1\zeta_3+3\zeta_3\zeta_2+95\zeta_3^2)g_4 - (4\zeta_1-\zeta_3)
     \right.\nonumber\\
 & \left. \times
      (16\zeta_1^2+4\zeta_1\zeta_2+44\zeta_1\zeta_3+5\zeta_3^2) \right] b^8g_h^4
\nonumber\\
& 
 + \left[-32(4\zeta_1+3\zeta_3)g_4 \right.\nonumber\\
 & \left. + (4\zeta_1-\zeta_3)(68\zeta_1+2\zeta_2+39\zeta_3) 
 \right] b^{10}g_h^{5} 
\nonumber\\
 & 
+ 64(\zeta_3-4\zeta_1) b^{12}g_h^6\,,\\
 c_{12} =&
  \frac{3}{4}\zeta_1^2g_4(\zeta_1\zeta_2+3\zeta_3^2)^2 \nonumber\\
 & 
  -\zeta_1(\zeta_1\zeta_2+3\zeta_3^2)(2\zeta_1\zeta_2+24\zeta_1\zeta_3+3\zeta_3^2)g_4b^2g_h
\nonumber\\
 &
  + \left[  \left( -4\zeta_1^4+10\zeta_1^3\zeta_2+10\zeta_1^3\zeta_3+\frac{5}{4}\zeta_1^2\zeta_2^2+\frac{1}{2}\zeta_1\zeta_3^2\zeta_2\right.\right.\nonumber\\
 & \left.\left.+\frac{115}{2}\zeta_1^2\zeta_3\zeta_2+\frac{855}{4}\zeta_1^2\zeta_3^2+127\zeta_1\zeta_3^3-7\zeta_3^4 \right) g_4 \right.\nonumber\\
 & \left.+ \frac{1}{4}\zeta_1^2(4\zeta_1-\zeta_3)(4\zeta_1^2-2\zeta_1\zeta_2-5\zeta_1\zeta_3-5\zeta_3^2) \right] b^4g_h^2
\nonumber\\
 &
  + \left[ \left(16\zeta_1^3-20\zeta_1^2\zeta_2-200\zeta_1^2\zeta_3-27\zeta_1\zeta_3\zeta_2 \right.\right. \nonumber\\
 & \left.\left. -415\zeta_1\zeta_3^2+3\zeta_3^2\zeta_2+8\zeta_3^3+4\frac{\zeta_3^4}{\zeta_1}\right)g_4 \right.\nonumber\\
 & \left. -\zeta_1(4\zeta_1-\zeta_3)(8\zeta_1^2-2\zeta_1\zeta_2-18\zeta_1\zeta_3-3\zeta_3^2)\right] b^6g_h^3 \nonumber\\
 & 
  + \left[ \left( -16\zeta_1^2+360\zeta_1\zeta_3+39\zeta_3^2-12\frac{\zeta_3^3}{\zeta_1}\right)g_4 \right.\nonumber\\
 & \left. +(4\zeta_1-\zeta_3)(20\zeta_1^2-2\zeta_1\zeta_2-57\zeta_1\zeta_3+3\zeta_3^2)  \right] b^8g_h^4 \nonumber\\
 &
-4 \left(4-\frac{\zeta_3}{\zeta_1}\right)(4\zeta_1^2-13\zeta_1\zeta_3+2\zeta_3^2)b^{10}g_h^5\,,\\
 c_{13} =&
\frac{3}{4}\zeta_1^2g_4(\zeta_1\zeta_2+3\zeta_3^2)^2\nonumber\\
 &
  -\zeta_1^2(\zeta_1\zeta_2+3\zeta_3^2)(\zeta_2+24\zeta_3)g_4b^2g_h
\nonumber\\
 &
  +\zeta_3(8\zeta_1-\zeta_3)(4\zeta_1\zeta_2+24\zeta_1\zeta_3+9\zeta_3^2)g_4b^4g_h^2
  \nonumber\\
 &
-4\frac{\zeta_3^2}{\zeta_1}(8\zeta_1-\zeta_3)^2g_4b^6g_h^3\,,\\
 c_{14} =&
\frac{1}{4}\zeta_1^2g_4(\zeta_1\zeta_2+3\zeta_3^2)^2\nonumber\\
 &
  -\zeta_1\zeta_3(\zeta_1\zeta_2+3\zeta_3^2)(8\zeta_1-\zeta_3)g_4b^2g_h
\nonumber\\
 &
+ \zeta_3^2(8\zeta_1-\zeta_3)^2g_4b^4g_h^2\,,\\
 c_{15} =&
  \frac{1}{8}\zeta_1^2\zeta_6(4\zeta_1-\zeta_3)b^2
  \nonumber\\
 &   + \frac{1}{2}(4\zeta_1^2-\zeta_1\zeta_2-5\zeta_1\zeta_3-2\zeta_3^2)
  \left[(\zeta_1-\zeta_3)g_4-\zeta_1^2\right]
\nonumber\\
 & 
  + \left[ -\frac{3}{8}\zeta_1\zeta_6(4\zeta_1-\zeta_3)b^2
\right.\nonumber\\
 & \left. + \left( 10\zeta_1^2-\zeta_1\zeta_2-\frac{5}{2}\zeta_1\zeta_3-\frac{1}{2}\zeta_3\zeta_2-12\zeta_3^2\right)g_4\right.\nonumber\\
 & \left.
      + \zeta_1\left(4\zeta_1^2-2\zeta_1\zeta_2-19\zeta_1\zeta_3-\frac{5}{2}\zeta_3^2\right)  \right] b^2g_h
\nonumber\\
 & 
+ \left[ \frac{1}{4}\zeta_6(4\zeta_1-\zeta_3)b^2
   + \left(-4\zeta_1+\frac{3}{2}\zeta_2+43\zeta_3\right)g_4 \right.\nonumber\\
 & \left.
   +26\zeta_1^2+\frac{5}{2}\zeta_1\zeta_2+\frac{71}{2}\zeta_1\zeta_3+\zeta_3^2\right]b^4g_h^2
\nonumber\\
 & 
  + (-68\zeta_1-48g_4-\zeta_2-15\zeta_3)b^6g_h^3 + 32b^8g_h^4\,,\\
 c_{16} =&
  \frac{1}{16}\zeta_1^2\zeta_6(4\zeta_1-\zeta_3)b^2 + \frac{1}{2}(4\zeta_1^2-\zeta_1\zeta_2-5\zeta_1\zeta_3-2\zeta_3^2)\nonumber\\
 & \times (\zeta_1-\zeta_3)g_4-\frac{1}{2}(4\zeta_1^2-\zeta_1\zeta_2-5\zeta_1\zeta_3-2\zeta_3^2)\zeta_1^2
\nonumber\\
 & 
+ \left[ -\frac{1}{8}\zeta_1\zeta_6(4\zeta_1-\zeta_3)b^2
   + \frac{1}{4}\left(12\zeta_1^2\right.\right.\nonumber\\
 & \left.\left.-2\zeta_1\zeta_2+29\zeta_1\zeta_3-\zeta_3\zeta_2-54\zeta_3^2+4\frac{\zeta_3^3}{\zeta_1}\right)g_4\right.\nonumber\\
 & \left. 
      +\frac{1}{4}\zeta_1(32\zeta_1^2-6\zeta_1\zeta_2-84\zeta_1\zeta_3-3\zeta_3^2)\right] b^2g_h
\nonumber\\
 & 
  + \left[ -\frac{1}{2\zeta_1}(28\zeta_1^2-55\zeta_1\zeta_3+6\zeta_3^2)g_4
		       \right.\nonumber\\
 & \left.
     +4\zeta_1^2+\zeta_1\zeta_2+37\zeta_1\zeta_3-\frac{7}{2}\zeta_3^2\right]b^4g_h^2
\nonumber\\
 & 
-2\left( 12\zeta_1+5\zeta_3-\frac{\zeta_3^2}{\zeta_1}\right)b^6g_h^3\,,\\
 c_{17} =&
  \frac{1}{2}(4\zeta_1^2-\zeta_1\zeta_2-5\zeta_1\zeta_3-2\zeta_3^2)(\zeta_1-\zeta_3)g_4\nonumber\\
 & - \frac{1}{2}(4\zeta_1^2-\zeta_1\zeta_2-5\zeta_1\zeta_3-2\zeta_3^2)\zeta_1^2
\nonumber\\
 & 
+ \left[ -\left(1-\frac{\zeta_3}{\zeta_1}\right) 
   (4\zeta_1^2-13\zeta_1\zeta_3+2\zeta_3^2)g_4 \right.\nonumber\\
 & \left. +\zeta_1(12\zeta_1^2-\zeta_1\zeta_2-23\zeta_1\zeta_3+\zeta_3^2)\right]b^2g_h
\nonumber\\
 & 
  + 2(-8\zeta_1^2+18\zeta_1\zeta_3-3\zeta_3^2)b^4g_h^2\,.
 \end{align}

\section{Coefficients of dispersion relation}
\label{app:calS}

In this appendix we show the coefficients $\mathcal{S}_{1,\cdots,11}$ of the dispersion relation (\ref{eqn:dispersionrelation})-(\ref{eqn:def-alphas}).
 \begin{align}
 \mathcal{S}_1 =& g_4\,,\\
 \mathcal{S}_2 =& \mathcal{N}_2-\mathcal{N}_1\,,\\
 \mathcal{S}_3 =& 
  -b^2 g_4 \mathcal{N}_1 \mathcal{N}_2 \mathcal{N}_3 \zeta_3^2-\mathcal{N}_1 \mathcal{N}_3 (\mathcal{N}_1^2 \mathcal{N}_2-\mathcal{N}_5) b^2 g_4 \nonumber\\
 & +[2 \mathcal{N}_3 \mathcal{N}_2 \mathcal{N}_1^2 b^2 g_4+\mathcal{N}_3 \mathcal{N}_2 \mathcal{N}_1^2 (\mathcal{N}_1-2 \mathcal{N}_2) b^2] \zeta_3 \nonumber\\
 & -\mathcal{N}_1^3 \mathcal{N}_2 \mathcal{N}_3 (\mathcal{N}_1-2 \mathcal{N}_2) b^2 \nonumber\\
 & -2 \mathcal{N}_2 \mathcal{N}_5 (\mathcal{N}_1+2 \mathcal{N}_2) (\mathcal{N}_1-\mathcal{N}_2)
  \,,\\
 \mathcal{S}_4 =&
  [-\mathcal{N}_1 \mathcal{N}_3 (2 \mathcal{N}_1-\mathcal{N}_2) b^2 g_4+4 (2 \mathcal{N}_1-\mathcal{N}_2)^2 (\mathcal{N}_1-\mathcal{N}_2)^2] \zeta_3^2 \nonumber\\
 & +[2 \mathcal{N}_1^2 (2 \mathcal{N}_1-\mathcal{N}_2) (\mathcal{N}_1-\mathcal{N}_2) b \zeta_4 \nonumber\\
 & +\mathcal{N}_3 \mathcal{N}_1^2 (7 \mathcal{N}_1-5 \mathcal{N}_2) b^2 g_4 \nonumber\\
 & -\mathcal{N}_3 \mathcal{N}_1^2 (3 \mathcal{N}_1^2-3 \mathcal{N}_1 \mathcal{N}_2+\mathcal{N}_2^2) b^2 \nonumber\\
& -8 \mathcal{N}_1 (2 \mathcal{N}_1-\mathcal{N}_2) (5 \mathcal{N}_1-4 \mathcal{N}_2) (\mathcal{N}_1-\mathcal{N}_2)^2] \zeta_3 \nonumber\\
& + b^2 \mathcal{N}_1^4 \zeta_4^2 /4 \nonumber\\
& +[- b^3 \mathcal{N}_1^4 \mathcal{N}_3 /2 -2 \mathcal{N}_1^3 (5 \mathcal{N}_1-4 \mathcal{N}_2) (\mathcal{N}_1-\mathcal{N}_2) b] \zeta_4 \nonumber\\
 & +b^2 \mathcal{N}_1^2 \mathcal{N}_5 \zeta_5-\mathcal{N}_1 \mathcal{N}_3 (5 \mathcal{N}_1^3-4 \mathcal{N}_1^2 \mathcal{N}_2-\mathcal{N}_5) b^2 g_4 \nonumber\\
 & + \mathcal{N}_1^3 \mathcal{N}_3 (15 \mathcal{N}_1^2-21 \mathcal{N}_1 \mathcal{N}_2+8 \mathcal{N}_2^2) b^2 /2 \nonumber\\
 & +4 (\mathcal{N}_1-\mathcal{N}_2)^2 (25 \mathcal{N}_1^4-40 \mathcal{N}_1^3 \mathcal{N}_2+16 \mathcal{N}_1^2 \mathcal{N}_2^2+\mathcal{N}_2 \mathcal{N}_5)\,,\\
 \mathcal{S}_5 =& 
  [-b^2 \mathcal{N}_1 \mathcal{N}_3 g_4^2+2 \mathcal{N}_2 (\mathcal{N}_1+2 \mathcal{N}_2) (\mathcal{N}_1-\mathcal{N}_2) g_4] \zeta_3^2 \nonumber\\
 & +\{ 2 \mathcal{N}_3 \mathcal{N}_1^2 b^2 g_4^2+[\mathcal{N}_3 \mathcal{N}_1^2 (\mathcal{N}_1-2 \mathcal{N}_2) b^2 \nonumber\\
 & -4 \mathcal{N}_1 \mathcal{N}_2 (\mathcal{N}_1+2 \mathcal{N}_2) (\mathcal{N}_1-\mathcal{N}_2)] g_4  \nonumber\\
  & -2 \mathcal{N}_1 \mathcal{N}_2 (\mathcal{N}_1+2 \mathcal{N}_2) (\mathcal{N}_1-\mathcal{N}_2) (\mathcal{N}_1-2 \mathcal{N}_2)\} \zeta_3 \nonumber\\
  & -b^2 g_4^2 \mathcal{N}_1^3 \mathcal{N}_3
  +[-\mathcal{N}_1^3 \mathcal{N}_3 (\mathcal{N}_1-2 \mathcal{N}_2) b^2 \nonumber\\
 & +2 (\mathcal{N}_1+2 \mathcal{N}_2) (\mathcal{N}_1-\mathcal{N}_2) (\mathcal{N}_1^2 \mathcal{N}_2-\mathcal{N}_5)] g_4 \nonumber\\
 & +2 \mathcal{N}_1^2 \mathcal{N}_2 (\mathcal{N}_1+2 \mathcal{N}_2) (\mathcal{N}_1-\mathcal{N}_2) (\mathcal{N}_1-2 \mathcal{N}_2)\,,\\
 \mathcal{S}_6 =& 
  [-\mathcal{N}_2 (2 \mathcal{N}_1-\mathcal{N}_2) b^2 g_4 \zeta_5-b^2 \mathcal{N}_1 \mathcal{N}_3 g_4^2] \zeta_3^2 \nonumber\\
 & +\{ 2 \mathcal{N}_1 \mathcal{N}_2 (\mathcal{N}_1-\mathcal{N}_2) b g_4 \zeta_4 \nonumber\\
 & +[\mathcal{N}_1 \mathcal{N}_2 (7 \mathcal{N}_1-5 \mathcal{N}_2) b^2 g_4 \nonumber\\
 & +\mathcal{N}_1 \mathcal{N}_2 (\mathcal{N}_1^2-3 \mathcal{N}_1 \mathcal{N}_2+\mathcal{N}_2^2) b^2] \zeta_5 \nonumber\\
 & +\mathcal{N}_1 \mathcal{N}_3 (5 \mathcal{N}_1-3 \mathcal{N}_2) b^2 g_4^2-\mathcal{N}_1 \mathcal{N}_2 \mathcal{N}_3 (2 \mathcal{N}_1-\mathcal{N}_2) b^2 g_4 \nonumber\\
 & -4 \mathcal{N}_1 \mathcal{N}_2^2 (\mathcal{N}_1-\mathcal{N}_2)^2\} \zeta_3+ b^2 g_4 \mathcal{N}_1^2 \mathcal{N}_2 \zeta_4^2 /4 \nonumber\\
 & +\{[- b^3 \mathcal{N}_1^2 \mathcal{N}_2 \mathcal{N}_3 /2 \nonumber\\
 & -\mathcal{N}_1 \mathcal{N}_2 (5 \mathcal{N}_1-3 \mathcal{N}_2) (\mathcal{N}_1-\mathcal{N}_2) b] g_4 \nonumber\\
 & -\mathcal{N}_1 \mathcal{N}_2^2 (\mathcal{N}_1-\mathcal{N}_2)^2 b\} \zeta_4 \nonumber\\
 & +[-\mathcal{N}_2 (5 \mathcal{N}_1^3-4 \mathcal{N}_1^2 \mathcal{N}_2-\mathcal{N}_5) b^2 g_4- b^4 \mathcal{N}_1^3 \mathcal{N}_2 \mathcal{N}_3 /4 \nonumber\\
 & - \mathcal{N}_1^2 \mathcal{N}_2 (5 \mathcal{N}_1^2-15 \mathcal{N}_1 \mathcal{N}_2+8 \mathcal{N}_2^2) b^2 /2 ] \zeta_5 \nonumber\\
 & - \mathcal{N}_1 \mathcal{N}_3 (5 \mathcal{N}_1-3 \mathcal{N}_2)^2 b^2 g_4^2 /4 \nonumber\\
 & + \mathcal{N}_1 \mathcal{N}_2 \mathcal{N}_3 (2 \mathcal{N}_1-\mathcal{N}_2) (5 \mathcal{N}_1-3 \mathcal{N}_2) b^2 g_4 /2 \nonumber\\
 & - \mathcal{N}_1 \mathcal{N}_2^2 \mathcal{N}_3 (\mathcal{N}_1-\mathcal{N}_2)^2 b^2 /4 \nonumber\\
 & +2 \mathcal{N}_1 \mathcal{N}_2^2 (5 \mathcal{N}_1-3 \mathcal{N}_2) (\mathcal{N}_1-\mathcal{N}_2)^2\,,\\
 \mathcal{S}_7 =&
  [- b^2 g_4 \mathcal{N}_1 \mathcal{N}_2 \zeta_5 /2 -b^2 \mathcal{N}_1 \mathcal{N}_3 g_4^2 \nonumber\\
 & +(\mathcal{N}_1-\mathcal{N}_2) (-\mathcal{N}_2+4 \mathcal{N}_1) (2 \mathcal{N}_1^2-3 \mathcal{N}_1 \mathcal{N}_2 \nonumber\\
 & +2 \mathcal{N}_2^2) g_4/\mathcal{N}_1] \zeta_3^2+\{\mathcal{N}_1 (2 \mathcal{N}_1-\mathcal{N}_2) (\mathcal{N}_1-\mathcal{N}_2) b g_4 \zeta_4 \nonumber\\
 & +[b^2 g_4 \mathcal{N}_1^2 \mathcal{N}_2+ \mathcal{N}_1^2 \mathcal{N}_2 (\mathcal{N}_1-2 \mathcal{N}_2) b^2/2] \zeta_5 \nonumber\\
 & + \mathcal{N}_3 \mathcal{N}_1 (7 \mathcal{N}_1-3 \mathcal{N}_2) b^2 g_4^2 /2 \nonumber\\
 & +[- \mathcal{N}_3 \mathcal{N}_1 (3 \mathcal{N}_1^2-4 \mathcal{N}_1 \mathcal{N}_2+3 \mathcal{N}_2^2) b^2 /2 \nonumber\\
 & -(\mathcal{N}_1-\mathcal{N}_2) (40 \mathcal{N}_1^3-85 \mathcal{N}_1^2 \mathcal{N}_2+73 \mathcal{N}_1 \mathcal{N}_2^2-22 \mathcal{N}_2^3)] g_4 \nonumber\\
 & -\mathcal{N}_2 (\mathcal{N}_1-\mathcal{N}_2) (\mathcal{N}_1^3-3 \mathcal{N}_1^2 \mathcal{N}_2+\mathcal{N}_1 \mathcal{N}_2^2-2 \mathcal{N}_2^3)\} \zeta_3 \nonumber\\
 & + b^2 g_4 \mathcal{N}_1^3 \zeta_4^2 /8+[- \mathcal{N}_1^3 \mathcal{N}_3 b^3 /4 \nonumber\\
 & -\mathcal{N}_1^2 (5 \mathcal{N}_1-4 \mathcal{N}_2) (\mathcal{N}_1-\mathcal{N}_2) b] g_4 \zeta_4 \nonumber\\
 & +[- \mathcal{N}_1 (\mathcal{N}_1^2 \mathcal{N}_2-\mathcal{N}_5) b^2 g_4  \nonumber\\
 & - \mathcal{N}_1^3 \mathcal{N}_2 (\mathcal{N}_1-2 \mathcal{N}_2) b^2 ] \zeta_5/2 \nonumber\\
 & - \mathcal{N}_1^2 \mathcal{N}_3 (5 \mathcal{N}_1-3 \mathcal{N}_2) b^2 g_4^2 /2 \nonumber\\
 & +[ \mathcal{N}_1^2 \mathcal{N}_3 (15 \mathcal{N}_1^2-23 \mathcal{N}_1 \mathcal{N}_2+12 \mathcal{N}_2^2) b^2 /4 \nonumber\\
 & +(\mathcal{N}_1-\mathcal{N}_2) (50 \mathcal{N}_1^5-125 \mathcal{N}_1^4 \mathcal{N}_2+116 \mathcal{N}_1^3 \mathcal{N}_2^2 \nonumber\\
 & -38 \mathcal{N}_1^2 \mathcal{N}_2^3-\mathcal{N}_1 \mathcal{N}_2 \mathcal{N}_5-2 \mathcal{N}_2^2 \mathcal{N}_5)/\mathcal{N}_1] g_4 \nonumber\\
 & + \mathcal{N}_1^2 \mathcal{N}_2 \mathcal{N}_3 (\mathcal{N}_1+2 \mathcal{N}_2) (\mathcal{N}_1-\mathcal{N}_2) b^2 /4 \nonumber\\
 & + \mathcal{N}_1 \mathcal{N}_2 (\mathcal{N}_1-\mathcal{N}_2) (5 \mathcal{N}_1^3-9 \mathcal{N}_1^2 \mathcal{N}_2 \nonumber\\
 & -10 \mathcal{N}_1 \mathcal{N}_2^2+8 \mathcal{N}_2^3) /2+\sqrt{\mathcal{S}_5 \mathcal{S}_6}\,,\\
 \mathcal{S}_8 =& \zeta_5\,,\\
 \mathcal{S}_9 =&
  -g_4 (\mathcal{N}_1-\zeta_3) (-\zeta_3 g_4+g_4 \mathcal{N}_1+\mathcal{N}_1^2-2 \mathcal{N}_1 \mathcal{N}_2)\,,\\
 \mathcal{S}_{10} =&
  -\zeta_3^2 g_4^2+[(5 \mathcal{N}_1-3 \mathcal{N}_2) g_4^2-g_4 \mathcal{N}_2^2] \zeta_3 \nonumber\\
 & - (5 \mathcal{N}_1-3 \mathcal{N}_2)^2 g_4^2 /4 +[- b^2 \mathcal{N}_1 \mathcal{N}_2 \mathcal{N}_3 /4 \nonumber\\
 & + \mathcal{N}_2^2 (5 \mathcal{N}_1-3 \mathcal{N}_2)/2] g_4  - \mathcal{N}_2^2 (\mathcal{N}_1-\mathcal{N}_2)^2 /4 \,,\\
 \mathcal{S}_{11} =& 
  \{-\zeta_3^2 g_4+[(7 \mathcal{N}_1 -3 \mathcal{N}_2) g_4 /2 + \mathcal{N}_1^2 /2\nonumber\\
 & -\mathcal{N}_1 \mathcal{N}_2- \mathcal{N}_2^2 /2 ] \zeta_3- \mathcal{N}_1 (5 \mathcal{N}_1-3 \mathcal{N}_2) g_4/2 \nonumber\\
 & - \mathcal{N}_3 \mathcal{N}_1^2 b^2/8 - \mathcal{N}_1(5 \mathcal{N}_1^2-13 \mathcal{N}_1 \mathcal{N}_2+4 \mathcal{N}_2^2)/4\} g_4  \nonumber\\
 & +\sqrt{\mathcal{S}_9 \mathcal{S}_{10}}\,.
 \end{align}

\section{Coefficients of the equation of motion in the long wavelength limit}
\label{app:calA123}

The coefficients $\mathcal{A}_{1,2,3}$ in (\ref{eqn:eom-tildeAodd-longwavelength}) are as follows.
\begin{align}
 \mathcal{A}_1 =& 4 (-\zeta_3+\zeta_1)^2\,,\\
 \mathcal{A}_2 =&
  (36 g_h + 7 \zeta_5/2) g_hb^4+9 b^3 \zeta_4 g_h/4 \nonumber\\
 &
  +[24(\zeta_1-\zeta_3) g_h+(3 \zeta_1- 6\zeta_3 - \zeta_2) \zeta_5/4- \zeta_4^2/16] b^2 \nonumber\\
&
  +3 \zeta_4 (-\zeta_3+\zeta_1) b/4+4 (-\zeta_3+\zeta_1)^2\,,\\
 \mathcal{A}_3 =&
  28 g_h^3 b^6/\zeta_1+(3 \zeta_1-2 \zeta_2-30 \zeta_3) g_h^2 b^4/\zeta_1 \nonumber\\
&
  -g_h \zeta_4 (-\zeta_3+\zeta_1) b^3/\zeta_1\nonumber\\
&
  +[(27 \zeta_1^2+(3 \zeta_2-30 \zeta_3) \zeta_1+12 \zeta_3^2) g_h/(2 \zeta_1) \nonumber\\
&
  +(-\zeta_3+\zeta_1)^2 \zeta_5/(2 \zeta_1)] b^2+3 \zeta_4 (-\zeta_3+\zeta_1) b/8 \nonumber\\
& +4 (-\zeta_3+\zeta_1)^2 + \sqrt{\mathcal{A}_1\mathcal{A}_2}\,.
\end{align}


\begin{thebibliography}{99}

\bibitem{Widrow:2002ud} 
  L.~M.~Widrow,
  ``Origin of galactic and extragalactic magnetic fields,''
  Rev.\ Mod.\ Phys.\  {\bf 74}, 775 (2002)
  doi:10.1103/RevModPhys.74.775

\bibitem{Kandus:2010nw} 
  A.~Kandus, K.~E.~Kunze and C.~G.~Tsagas,
  ``Primordial magnetogenesis,''
  Phys.\ Rept.\  {\bf 505}, 1 (2011)
  doi:10.1016/j.physrep.2011.03.001
	
\bibitem{Durrer:2013pga} 
  R.~Durrer and A.~Neronov,
  ``Cosmological Magnetic Fields: Their Generation, Evolution and Observation,''
  Astron.\ Astrophys.\ Rev.\  {\bf 21}, 62 (2013)
  doi:10.1007/s00159-013-0062-7
	
\bibitem{Subramanian:2015lua} 
  K.~Subramanian,
  ``The origin, evolution and signatures of primordial magnetic fields,''
  Rept.\ Prog.\ Phys.\  {\bf 79}, no. 7, 076901 (2016)
  doi:10.1088/0034-4885/79/7/076901

\bibitem{Mukohyama:2016npi} 
  S.~Mukohyama,
  Phys.\ Rev.\ D {\bf 94}, no. 12, 121302 (2016)
  doi:10.1103/PhysRevD.94.121302
  [arXiv:1607.07041 [hep-th]].

\bibitem{Horndeski:1974wa} 
  G.~W.~Horndeski,
  ``Second-order scalar-tensor field equations in a four-dimensional space,''
  Int.\ J.\ Theor.\ Phys.\  {\bf 10}, 363 (1974).
  doi:10.1007/BF01807638

\bibitem{Deffayet:2011gz} 
  C.~Deffayet, X.~Gao, D.~A.~Steer and G.~Zahariade,
  ``From k-essence to generalised Galileons,''
  Phys.\ Rev.\ D {\bf 84}, 064039 (2011)
  doi:10.1103/PhysRevD.84.064039

\bibitem{Horndeski:1976gi} 
  G.~W.~Horndeski,
  ``Conservation of Charge and the Einstein-Maxwell Field Equations,''
  J.\ Math.\ Phys.\  {\bf 17}, 1980 (1976).
  doi:10.1063/1.522837

\bibitem{Gumrukcuoglu:2016jbh} 
  A.~E.~Gumrukcuoglu, S.~Mukohyama and T.~P.~Sotiriou,
  ``Low energy ghosts and the Jeans’ instability,''
  Phys.\ Rev.\ D {\bf 94}, no. 6, 064001 (2016)
  doi:10.1103/PhysRevD.94.064001
  [arXiv:1606.00618 [hep-th]].

\bibitem{TheLIGOScientific:2017qsa} 
  B.~P.~Abbott {\it et al.} [LIGO Scientific and Virgo Collaborations],
  Phys.\ Rev.\ Lett.\  {\bf 119}, no. 16, 161101 (2017)
  doi:10.1103/PhysRevLett.119.161101
  [arXiv:1710.05832 [gr-qc]].

\bibitem{Monitor:2017mdv} 
  B.~P.~Abbott {\it et al.} [LIGO Scientific and Virgo and Fermi-GBM and INTEGRAL Collaborations],
  Astrophys.\ J.\  {\bf 848}, no. 2, L13 (2017)
  doi:10.3847/2041-8213/aa920c
  [arXiv:1710.05834 [astro-ph.HE]].

 \bibitem{nextproject}
 R.~Durrer, Y.~Huang, T.~Kahniashvili, S.~Mandal and S.~Mukohyama, work in progress. 

\end{thebibliography}
\end{document}